\documentclass[10pt]{JHEP3}

\usepackage{amsmath}
\usepackage{amsfonts}
\usepackage{amssymb}
\usepackage{graphicx}
\usepackage{epsfig}
\usepackage{hhline}

\newcommand{\Tr}{\mathrm{Tr}}

\newcommand{\beq}{\begin{eqnarray}}
\newcommand{\eeq}{\end{eqnarray}}
\newcommand{\ket}[1]{\left| #1 \right\rangle}
\newcommand{\bra}[1]{\left\langle #1 \right|}

\newcommand{\VEV}[1]{\left\langle #1 \right\rangle}
\newcommand{\RC}{\mathcal R}
\preprint{WUB/10-19}
\title{Glueball masses in the large $N$ limit}

\author{Biagio Lucini \\
  School of Physical Sciences, Swansea University\\
  Singleton Park, Swansea SA2 8PP, UK\\
  E-mail: \email{b.lucini@swansea.ac.uk} 
}
\author{Antonio Rago\\
  Department of Physics, Bergische Universit\"at Wuppertal\\
  Gaussstr. 20, D-42119 Wuppertal, Germany\\
  E-mail: \email{rago@physik.uni-wuppertal.de} 
}
\author{Enrico Rinaldi\\
  SUPA, School of Physics and Astronomy, University of Edinburgh\\
  Edinburgh EH9 3JZ, UK\\
  E-mail: \email{e.rinaldi@sms.ed.ac.uk} 
}

\abstract
{ The lowest-lying glueball masses are computed in SU($N$) gauge theory 
  on a spacetime lattice for constant value of the lattice spacing $a$
  and for $N$ ranging from $3$ to $8$. The lattice spacing is fixed
  using the deconfinement temperature at temporal extension of the
  lattice $N_T = 6$. The calculation is conducted employing in each
  channel a variational ansatz performed on a large basis of operators
  that includes also torelon and (for the lightest states) scattering
  trial functions. This basis is constructed using an automatic algorithm that
  allows us to build operators of any size and shape in any irreducible
  representation of the cubic group. A good signal is extracted for
  the ground state and the first excitation in several symmetry channels. It is shown that all the observed states are well
  described by their large $N$ values, with modest ${\cal O}(1/N^2)$
  corrections. In addition spurious states are identified that couple to torelon and scattering operators.
  As a byproduct of our calculation, the critical couplings for the
  deconfinement phase transition for $N=5$ and $N=7$ and temporal
  extension of the lattice $N_T=6$ are determined.}


\begin{document}
\section{Introduction}
\label{sect:introduction}
Non-Abelian SU($N$) gauge theories with fermions in the fundamental representation have a simpler diagrammatic expansion in the limit in which the number of colours $N$ goes to infinity with the product $\lambda = g^2 N$ held fixed~\cite{Hooft:1973jz,Veneziano:1974ag,Witten:1979kh} (see e.g.~\cite{Manohar:1998xv,Makeenko:1999hq} for reviews of the original ideas and some more recent developments). Due to their simpler structure that can nevertheless shed invaluable insights on non-perturbative phenomena in non-Abelian gauge theories, SU($N$) gauge theories in the large $N$ limit play a central role in the gauge-gravity correspondence and have become the subject of a line of numerical investigations on the lattice. In addition to determining values for observables in the large $N$ limit, lattice calculations provide their corrections at finite $N$, which can be expressed as a power series in $1/N^2$ for the quenched theory and in $1/N$ in the dynamical case; the emerging picture is that, at least for the quenched theory, only the leading correction of ${\cal O}(1/N^2)$ is sufficient to describe the system at any finite value of $N$ bigger than two at a level of accuracy of the order of a few percents. These calculations clarify quantitatively the meaning of the statement that the physical case $N = 3$ is close to $N = \infty$. Numerical calculations at large $N$ have been recently reviewed in~\cite{Teper:2008yi,Narayanan:2009xh,Teper:2009uf}, with a  review of earlier numerical works provided in~\cite{Das:1984nb}.

If the lattice legitimates the use of the large $N$ limit to describe the physics for $N=3$, analytical approaches based on large $N$ ideas are of great help to model the lattice data. In order to assess the reliability of various analytical methods based on the large $N$ framework (which often have to resort to other approximations in addition to taking the large $N$ limit), it is important to compare their predictions to the lattice data for observables that are well under control in both approaches. The glueball spectrum in the pure Yang-Mills theory is one of the easiest observables to compare. Analytical results for glueball spectra in the large $N$ limit are available from calculations based on different approaches, in particular in various backgrounds in the gauge/string duality framework~\cite{Csaki:1998qr,Ooguri:1998hq,Brower:1999nj,Apreda:2003sy,Caceres:2000qe,Caceres:2005yx,Berg:2005pd,Berg:2006xy,Elander:2009pk}, with a variational ansatz for the groundstate of the Yang-Mills theory~\cite{Leigh:2005dg,Leigh:2006vg,Freidel:2006qy} and in the light-cone framework~\cite{Dalley:1998qa,Dalley:1999ii,Dalley:2000ye}. On the lattice side, glueballs in the SU(2) and mostly SU(3) Yang-Mills theory have been the subject of extended studies for quite a long time~\cite{Ishikawa:1983xg,Ishikawa:1983js,Berg:1982kp,Berg:1983qd,Berg:1986cj,Michael:1988jr,Bali:1993fb}, and recent investigations have provided us with reliable determinations of the spectrum in all the $J^{PC}$ channels~\cite{Morningstar:1997ff,Chen:2005mg}. Reliable numerical results at larger $N$ have become available only more recently~\cite{Teper:1998te,Teper:1998kw,Lucini:2001ej,Lucini:2002wg,Lucini:2004my,Lucini:2004eq,Meyer:2004jc,Meyer:2004vr,Meyer:2004gx} and mostly concern the groundstate and the first excitation in the $0^{++}$ and $2^{++}$ channels. 

In general, hadronic states on the lattice are computed with a variational ansatz, and in order to reliably extract the energy of excitations the variational basis must be large. Building a large variational basis in all the symmetry channels, corresponding to the quantum numbers of the hadrons for which we want to extract the mass spectrum, proves to be quite complicated from a technical point of view (see e.g. Refs.~\cite{Foley:2010te,Dudek:2010wm} for a discussion in the context of the meson spectrum). For this reason, so far the bulk of the effort for extracting reliable single-particle spectra in the glueball sector has been put on the physical case $N = 3$, with the calculations at larger $N$ focusing mostly on much fewer states. Given the theoretical importance of determining accurately the behaviour of the theory at large $N$, it would be useful to have a more comprehensive calculation also in that case.

Our work aims to fill this gap. In this paper we will provide the first determination of the large $N$ glueball spectrum (obtained with an extrapolation including values of $N$ up to eight) in several irreducible representations of the lattice rotational group and for both values of parity and charge conjugation. We find that, also for the states studied for the first time in this work, a small $1/N^2$ correction to the $N=\infty$ limit describes the data with an accuracy of a few percents all the way down to $N=3$. Progress over previous computations has been made possible by the implementation of an automatic method for generating computer code for correlators of operators in various representations of the lattice rotational group starting from some basic operators. With this technique, inserting extra operators in our variational basis becomes an almost trivial task and the only limitations to the number of operators that can be used in the calculation are given by the available computational resources. In particular, using our method we are able to build a variational basis that includes torelon (i.e. states coupling to two Polyakov loops, which have an infinite mass in the infinite volume limit) and scattering operators; a variational ansatz on such a comprehensive basis allows us to disentangle genuine single-particle states from spurious or multi-particle resonances. The investigation reported in this work has been performed at a single value of the lattice spacing that has been fixed to a common value across the various SU($N$) gauge groups simulating at the gauge coupling corresponding to the deconfinement temperature when the temporal extension of the lattice is $N_T=6$. The extrapolation to the continuum limit will be the subject of future investigations.

The rest of the paper is organised as follows. The lattice setup will be briefly discussed in Sect.~\ref{sect:lattice}. In Sect.~\ref{sect:group_operators} we shall review the variational method for extracting glueball masses, discuss scattering and torelon states, and briefly present our algorithm for automatically generating glueball operators in an irreducible representation of the rotational group. Our numerical results will be discussed in Sect.~\ref{sect:numerical}, with the extrapolation to large $N$ given in Sect.~\ref{sect:lnspectrum}. Finally, a summary of our findings together with an outlook on future developments will be the subject of Sect.~\ref{sect:conclusions}.

\section{SU($N$) gauge theories on the lattice}
\label{sect:lattice}
The lattice discretisation of SU($N$) Yang-Mills theory used throughout this work is entirely conventional. We consider the system defined on an isotropic four-dimensional torus of linear size $L$. If $a$ is the lattice spacing, the number of points in each direction is given by $N_L = L/a$. The volume of the system in physical units is given by $V = L^4$. Directions are indicated with Greek symbols and lattice points are labelled with Latin letters. The fundamental degrees of freedom of the theory are the link variables $U_{\mu}(i) \in$ SU($N$). $U_{\mu}(i)$ is associated to the bond $(i;\hat{\mu})$ that joins the site $i$ with $i + \hat{\mu}$. The Wilson action for the lattice theory is given by
\beq
S = \beta \sum_{i,\mu > \nu} \left( 1 - \frac{1}{N} \mbox{Re} \ \Tr \left( U_{\mu \nu}(i) \right) \right) \ ,
\eeq 
where $U_{\mu \nu}(i)$ is the parallel transport of the link variables along the elementary lattice plaquette stemming from point $i$ and identified by the positive pair of directions $(\hat{\mu},\hat{\nu})$. $\beta$ is defined as $\beta = 2N/g_0^2$, with $g_0$ the bare gauge coupling. The Euclidean path integral reads
\beq
Z = \int \left( {\cal D} U\right) e^{- S} \ ,
\eeq
with ${\cal D} U$ the product of the Haar measures associated to each link.

The observables we shall study in this work are glueball masses. For the continuum theory, at fixed quantum numbers $J^{PC}$ and for fixed excitation in the spectrum, the corresponding mass has corrections that can be expressed in a power series in $1/N^2$. This is also true in the lattice strong coupling~\cite{Hooft:2002yn} and has been shown to apply also for intermediate couplings~\cite{Lucini:2004my,Lucini:2004yh,Lucini:2005vg}. In this work, we study the large $N$ limit for fixed lattice spacing $a$ as $N$ varies. The lattice spacing is chosen in the domain where the behaviour of the discretised theory is dominated by the physics in the continuum limit, i.e. in a regime in which lattice corrections to continuum values of observables are modest.

Since like QCD SU($N$) gauge theories are defined by one dynamically generated scale, the physical value of the lattice spacing can be determined by measuring this length in units of $a$ and assigning to the dynamical scale its continuum value. Conventionally, the square root of the string tension $\sigma$ is chosen to set the scale; fixing another quantity as a function of $a$ amounts to corrections of order $a^2 \sigma$ to physical observables. In order to compare quantities at fixed lattice spacing across different SU($N$) groups, it proves useful to set the scale using the (pseudo)--critical coupling at fixed temporal extent. The justification for this choice (already successfully used in~\cite{DelDebbio:2007wk,Armoni:2008nq}) is that this quantity can be determined to a very high degree of accuracy. We adopt the deconfinement temperature at $N_T = 6$ to fix the lattice spacing to the same value when changing $N$. The choice of using $\beta_c$ at this particular value of $N_T$ to set the scale is a compromise between the requirement of being in the continuum scaling regime and the practical convenience of having sufficiently large lattices in physical units for a computationally bearable value of the number of lattice sites $N_L^4$. In fact, $N_L = 12$ for $\beta = \beta_c(N_T = 6)$ gives a glueball spectrum in the scaling region and free from large finite size artefacts~\cite{Lucini:2004my}.  
\section{Extracting glueball masses}
\label{sect:group_operators}
In this section we present the construction of our operators and we review
the general methodology for extracting glueball masses. While the standard variational procedure and
the construction of operators in irreducible representations of the cubic lattice group is well known,
this is, up to our knowledge, the first systematic attempt of inserting scattering and torelon
operators into the variational set, in order to rule out from the spectrum contributions of these spurious states.

\subsection*{Symmetries of the lattice spectrum}
At finite lattice spacing $a$ the continuum rotation group is not an exact symmetry of the system. 
The full continuum rotational symmetry  is dynamically restored only when $a \rightarrow 0$.
On the lattice, eigenstates of the Hamiltonian have to fall into the irreducible representations
of the octahedral point group $\mathcal{G}_O$, the symmetry group of the cube. 
The octahedral point group has 5 irreducible representations $A_1$, $A_2$, $E$, $T_1$ and $T_2$ respectively
with dimensions $1$, $1$, $2$, $3$, $3$. \\
Since we are interested in the glueball spectrum of the gauge theory in the continuum, we need to 
consider $\mathcal{G}_O$ as a subgroup of the complete rotation group SO($3$): irreducible representations of
SO($3$) are decomposed in terms of those of $\mathcal{G}_O$. 
Irreducible representations of integer spin $J$ in SO($3$) restricted to $\mathcal{G}_O$
are referred to as \emph{subduced} representations $J \downarrow \mathcal{G}_O$.
When considered as a representation of $\mathcal{G}_O$, the ($2J+1$) degeneracy of the continuum spin $J$ state is split
onto different irreducible representations of $\mathcal{G}_O$. 
A simple example of this kind of pattern is the spin $2$ (tensor) glueball, whose $5$ polarisations are seen on the lattice as 
different states, $2$ in the $E$ and $3$ in the $T_2$ representation of $\mathcal{G}_O$. 
Due to the breaking of continuum rotational symmetry on the lattice,
the aforementioned pattern of degeneracies is exact in the limit $a \to 0$,
but it is only approximate at finite $a$. Comparing the measured glueball spectrum with the expected pattern of degeneracy  can give information on the relevance of lattice artifacts.\\
Near the continuum limit, it is possible to identify the masses of spin $J$ glueballs by matching the patterns of degeneracies of the subduced representations $J \downarrow \mathcal{G}_O$
from the degeneracy coefficients. We report these coefficients up to $J=4$ in Tab.~\ref{tab:subd-reps}.
\TABLE[ht]
{%\begin{table}[ht]
%\begin{center}
  \begin{tabular}[h]{c|ccccc}
% \hline
    $J$ & $A_1$ & $A_2$ & $E$ & $T_1$ & $T_2$ \\
    \hline
    0 & 1 & 0 & 0 & 0 & 0 \\
    1 & 0 & 0 & 0 & 1 & 0 \\
    2 & 0 & 0 & 1 & 0 & 1 \\
    3 & 0 & 1 & 0 & 1 & 1 \\
    4 & 1 & 0 & 1 & 1 & 1 \\
% \hline
  \end{tabular}
%\end{center}
\caption{Subduced representations $J \downarrow \mathcal{G}_O$ of the octahedral group up to $J=4$.
   This table illustrates the spin content of the irreducible
   representations of $\mathcal{G}_O$ in terms of the continuum $J$.}
\label{tab:subd-reps}
}
%\end{table}\\
For any given operator $\bar{\mathcal{O}}$ on the lattice, we define a rotation 
transformation as $\RC_i(\bar{\mathcal{O}})$ where the index $i$ labels all the elements of the group $\mathcal{G}_O$.\\
Since a generic representation of the group will not be irreducible, in order to create states that transform only in a given symmetry channel, we will need to create an appropriate linear combination of the rotations of the original operator.
We define an operator in the irreducible representation $R$ as
\begin{equation}
  \label{eq:linear-combination}
  \Phi^{(R)}(t) \; = \;  \sum_i c_i^{(R)} \RC_i(\bar{\mathcal{O}}(t))
\quad .
\end{equation}
The coefficients $c_i^{(R)}$ appearing in the summation are obtained
from the unitary operator that implements the change of basis from our
choice of the representation in $24$ dimensions into an orthonormal basis for each of the $5$ invariant subspaces~\cite{Michael:1988jr}.\\
By adding parity and charge conjugation to the group of pure rotations, we get the full symmetry
group of glueball states on the lattice, which is referred to as $\mathcal{G}_O^{PC}$. The group $\mathcal{G}_O^{PC}$ has a total of 20 irreducible representations
labelled by $R^{PC}$, where $R$ indicates one of the $5$ irreducible
representations of $\mathcal{G}_O$, $P$ is the parity eigenvalue and
$C$ is the charge conjugation eigenvalue. 
As we will explain in details below, operators that are eigenstates of
irreducible representations of $\mathcal{G}_O$ are constructed as
traces of products of link variables along closed paths. Hence,
eigenstates of $C$ are given by the real ($C=+1$) or the imaginary ($C=-1$) part of the operator and adding and subtracting parity transformed operators gives definite--$P$ states (respectively $P=+1$ and $P=-1$).\\
As usual we will label as $R^{PC}$ the ground state in a given symmetry channel, and its excitations as $R^{PC}$ followed by one or more $^\star$.

\FIGURE[ht]{
\epsfig{file=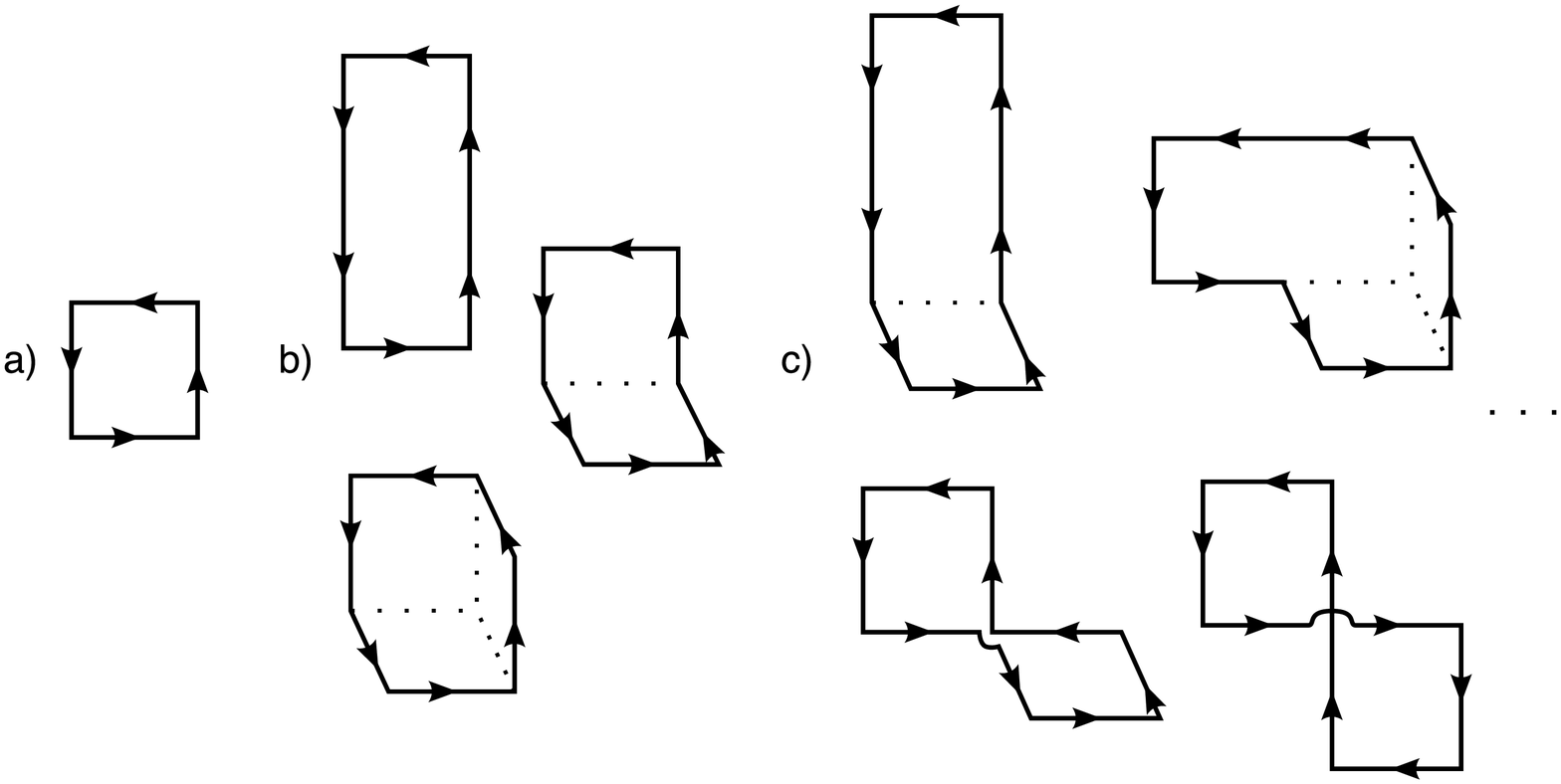,width=0.61\textwidth}
\label{fig:proto-path}
\caption{Set of basic prototypical paths used to construct operators in all the $20$ $R^{PC}$ symmetry channels.}
}

\subsection*{Construction of the operators}
At finite volume, the single-particle glueball spectrum receives
non-negligible corrections from multi--glueball states. Moreover, when the
system is closed with periodic boundary conditions (like in our case)
topological excitations wrapping the compact direction ({\em torelons}) with
the same quantum numbers of glueballs appear; if not correctly accounted for,
these states can affect significantly the measured glueball spectrum. 
In order to control these spurious contributions, we included in the
variational set operators that best overlap with two--glueballs and torelon
states.\\
The variational procedure described in the next section will allow us not only
to extract some excited states together with the lowest--lying state in each
channel, but also to disentangle the contributions of spurious states.\\
Before discussing the details of the variational procedure, we will describe
the kind of operators used to define the variational set.\\
All the operators described in the following are to be understood as gauge--invariant, vacuum--subtracted operators:
\begin{equation}
  \label{eq:op-general}
  \bar{\mathcal{O}}(t) \; = \; \mathcal{O}(t) - \bra{0}\mathcal{O}(t)\ket{0}
 \quad .
\end{equation}

\subsubsection*{Glueball operators}
\TABLE[h]{
%\begin{table}[ht!]
%\centering
    \begin{tabular}[h]{c|cccc}
       & ${++}$ & ${-+}$ & ${+-}$ & ${--}$ \\
      \hline
      $A_1$& 8 & 2 & 1  & 3 \\
      $A_2$& 3  & 1  & 3  & 3\\
       $E$ & 22  & 7  & 7   & 14\\
      $T_1$& 19 & 24 & 48 & 27 \\
      $T_2$& 44 & 33 & 33 & 29 \\
    \end{tabular}
   \caption{Number of different glueball operators calculated in each of the $20$ symmetry channels.}
  \label{tab:op-summary}
}%\end{table}

The single--trace operator that we use to project onto glueball states is simply defined as
\begin{equation}
  \label{eq:glueb-op}
  \mathcal{O}(t) \; = \; \phi(t)
  \quad ,
\end{equation}
where $\phi(t)$ is a zero-momentum operator given by the wall average over the temporal slice
\begin{equation}
  \label{eq:single-trace}
  \phi(t) \; = \; \frac{1}{N_L^3} \sum_{x\in\Lambda_s} \phi(x,t)
  \hspace{1cm} \phi(x,t)=\Tr \prod_{(i;\hat{\mu}) \in\cal{C}}U_{\mu}(i)
  \quad .
\end{equation}
In our definition of the variational set we used a wide range of different
closed loops ${\cal C}$, with lengths ranging from $4$ to $8$ lattice spacings. 
The basic shapes we used in our calculations are summarised in Fig.~\ref{fig:proto-path}.\\
In a given symmetry channel, we then correlate glueball operators of the form
\begin{equation}
\label{eq:glue-op-repr}
\begin{split}
\Phi^{(R)}(t) \; = \;  \sum_i c_i^{(R)} \RC_i(\phi(t)) -  \sum_i c_i^{(R)} \RC_i(\VEV{\phi(t)})\\
\; = \; \sum_i c_i^{(R)} \RC_i(\phi(t)) -  \VEV{\phi(t)} \sum_i c_i^{(R)} 
\quad , 
\end{split}
\end{equation}
where the last term in second line is different from zero only when
$R=A_1^{++}$. The number of different operators that we included in
our variational set is summarised in Tab.~\ref{tab:op-summary} for all
the symmetry channels.

\subsubsection*{Scattering operators}

An operator that projects onto scattering states of two glueballs is a double--trace operator. 
A trial operator that overlaps mainly onto scattering states has the form
\begin{equation}
  \label{eq:double-trace}
  \mathcal{O}(t) \; = \; (\phi(t) - \VEV{\phi(t)})^2
  \quad ,
\end{equation}
where $\phi(t)$ is a zero--momentum operator defined as in Eq.~(\ref{eq:single-trace}). 
The operator appearing in the correlator matrix for the scattering states can be written using
Eq.~(\ref{eq:linear-combination}) as
\begin{equation}
\label{eq:scatt-op-repr}
\begin{split}
\Phi^{(R)}(t) \; = \;  \sum_i c_i^{(R)} \RC_i\Big(\left(\phi(t) - \VEV{\phi(t)}\right)^2\Big) - 
                      \sum_i c_i^{(R)} \RC_i\Big(\VEV{(\phi(t) - \VEV{\phi(t)})^2}\Big)\\
              \equiv  \sum_i c_i^{(R)} \RC_i\left(\left(\phi(t) - \VEV{\phi(t)}\right)^2\right) - 
                      \left(\VEV{\phi^2(t)} - \VEV{\phi(t)}^2\right) \sum_i c_i^{(R)}
\quad . 
\end{split}
\end{equation}
As in the previous case the last term vanishes for $R \neq A_1^{++}$,
but the local subtraction of $\VEV{\phi(t)}$ in the first term will
appear in all the representations and is crucial in order to obtain
the correct two--point function, even though $\VEV{\phi(t)}$
alone, once appropriately symmetrised, would be different from zero only in the $A_1^{++}$ channel.\\ 
The local operators $\phi(x,t)$ have been chosen among the closed loops in Fig.~\ref{fig:proto-path} to create a variational
set for the scattering states as big as the one used for glueball states.\\
A technical aside: the aforementioned local subtraction in the scattering operators leads to
the necessity of having access to the values of their
vacuum expectation value ($VEV$) during the simulation, which implies
that the $VEV$s must be calculated prior to the evaluation of the scattering
operators. Moreover the $VEV$ values have to be known with an accuracy
higher than the most precise scattering operator. 

\FIGURE[hbt]{
\epsfig{figure=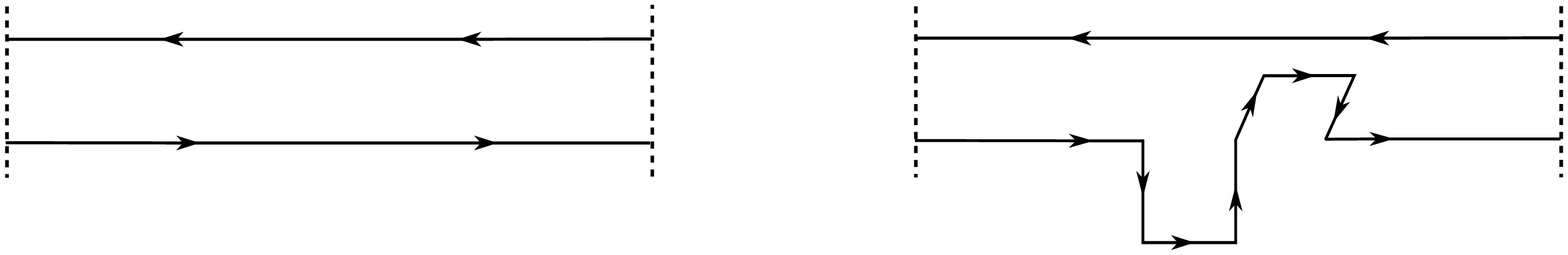,width=0.61\textwidth}
\label{fig:tor-path}
\caption{Paths used for the construction of operators coupling with
  torelon states. Periodic Boundary Conditions apply at the edges represented by the dashed lines.}
}

\subsubsection*{Torelon operators}
\TABLE[tb]{
%\begin{table}[h]
%\centering
  \begin{tabular}{c|ccccc}
       & ${++}$ & ${-+}$ & ${+-}$ & ${--}$ & ~ \\
      \hline
      $A_1$& 2 & 1 & 0  &  0 \\
      $A_2$& 1 & 0 & 1  &  1 \\
       $E$ & 7 & 3 & 3  &  3 \\
      $T_1$& 3 & 3 & 14 &  9 \\
      $T_2$& 9 & 9 & 8  &  3 \\
    \end{tabular}
   \caption{Number of different torelon operators calculated in each of the $20$ symmetry channels.}
  \label{tab:tor-summary}
%\end{table}
}A torelon state is an excitation winding around the toroidal lattice in the spatial direction.
The mass of a torelon state scales linearly with the lattice size $L$ and
the contribution of these states in the spectrum can be easily
identified with a finite volume study. However, we can include in our variational set operators that mainly project onto torelon states and identify them on a single volume.\\
Torelon excitations transform non-trivially under the centre of
the gauge group, and are characterised by their charge under this
transformation ($n$--ality). Since the glueballs transform
trivially under the centre of the gauge group, they can only couple to
states that have zero $n$--ality. For this reason the torelon
operators have been created from products of two Polyakov loops
$l_\nu$ winding around opposite directions. We started from defining
\begin{equation}
  \label{eq:torelon-op}
  \mathcal{O}(t) \; = \; \frac{1}{2N_L^2} \sum_{\mu \ne \nu} \sum_x l_\nu(x,t)l_\nu^{\dag} (x+\hat{\mu} a,t)
  \quad ,
\end{equation}
where the sum over $\mu$ runs on the spatial directions orthogonal to
the one of the loops, and then the operator $\Phi^{(R)}(t)$ can be
constructed as
in the previous cases by using Eq.~(\ref{eq:linear-combination}).\\
By choosing different shapes for the combination $l_\nu(x,t)l_\nu^{\dag} (x+\hat{\mu} a,t)$, we can obtain a fairly large variational
set in all the $20$ symmetry channels (see Tab.~\ref{tab:tor-summary}). The shapes we used are pictorially shown in Fig.~\ref{fig:tor-path}.

\subsection*{Blocking and smearing}
Masses of a lattice state $R^{PC}$ can be extracted from the behaviour of the correlator 
\begin{equation}
  \begin{split}
  \label{eq:corr-easy}
  C^{(R)}(t) \; &= \; \bra{0} \Phi^{(R)\dag}(t) \Phi^{(R)}(0) \ket{0}\\
  \; &= \; \sum_G \left| \bra{G} \Phi^{(R)} \ket{0} \right|^2 
  \exp{(-m_G^{(R)} t)}
\quad ,
\end{split}
\end{equation}
where $G$ labels a generic eigenstate of the action in an appropriate
orthonormal basis. At large temporal distance, when only the lowest--lying state, labelled by the subscript $0$, is contributing to the sum, it becomes possible to determine the 
mass of this state from the asymptotic single exponential decay
\begin{equation}
  \label{eq:corr-large-t}
  \lim_{t \rightarrow \infty} C^{(R)}(t) \; = \; \left| c_0 \right|^2 \exp{(-m_0^{(R)}
    t)}
  \quad ,
\end{equation}
where we have defined the overlap of the operator $\Phi^{(R)}$ on the state $G$ as
\begin{equation}
  \label{eq:overlap-corr}
  \left| c_G \right|^2 \; = \; \left| \bra{G} \Phi^{(R)} \ket{0}
  \right|^2
\end{equation}
and the normalisation is chosen in such a way that
\begin{equation}
  \label{eq:overlap-norm}
  \sum_G  \left| c_G \right|^2 \; = 1
 \quad .
\end{equation}

In principle, Eq.~(\ref{eq:corr-large-t}) allows one to extract glueball
masses from the large $t$ behaviour of correlators. However, it is already well
known~\cite{Albanese:1987ds,Teper:1987wt,Michael:1988jr,Morningstar:1997ff}
that a good signal--over--noise ratio
for the decay of this gluonic correlator can be obtained only at small temporal
distances. Therefore it is mandatory to obtain the asymptotic
single--exponential behaviour for $t$ as small as possible.
In fact, the single-exponential behaviour could be obtained at any $t$ if there were only a single propagating state.
The main idea underlying the variational technique is to enhance the propagation of a single given state $G$ in the sum of
Eq.~(\ref{eq:corr-easy}) and be able to find a normalised
overlap $\left| c_G \right|^2$ of order $1$. To achieve such a result, standard smearing and blocking
techniques~\cite{Albanese:1987ds,Teper:1987wt}
have been implemented in the past, together with variational
procedures.

In this work, we used the
improved blocking algorithm of~\cite{Lucini:2004my}. The basic idea is
to create a configuration of super--links, obtained from an appropriate
composition of blocking and smearing, from which we then build our ``blocked'' operators.
The operator resulting from this procedure has a typical ``size'' given by the level
of iterations of the algorithm. Blocked operators have a smaller
sensitivity to fluctuations at the UV scale, and their overlap with the
high--energy modes of the spectrum is reduced.\\
In using the algorithm of~\cite{Lucini:2004my}, we fixed the parameters of the algorithm to $(p_a,p_d)=(0.40,0.16)$ for the smearing part (see Fig.~\ref{fig:smear-path}), while a blocked link is obtained multiplying two consecutive smeared links. This choice has been shown to produce operators with $\left| c_G \right|^2 \approx O(1)$ for all symmetry channels. \\
As a result, operators constructed at different blocking levels $N_b$
will allow us to enlarge the variational set 
introduced above. For example, each of the glueball operators
of Tab.~\ref{tab:op-summary} can be built using blocked super--links
at different levels and, in our variational procedure, we were thus
able to use a set that is effectively $N_b=4$ times bigger than the original one.\\
A remark is in order here. The blocking procedure may lead to the construction of gauge variant operators
if the size of the lattice along the blocked direction is not multiple
of $2^{N_b-1}$ and one is measuring operators that wind along this
direction. This problem can be circumvented mixing different lower blocking levels~\cite{Teper:1998te}.

\FIGURE[ht]{
  \epsfig{file=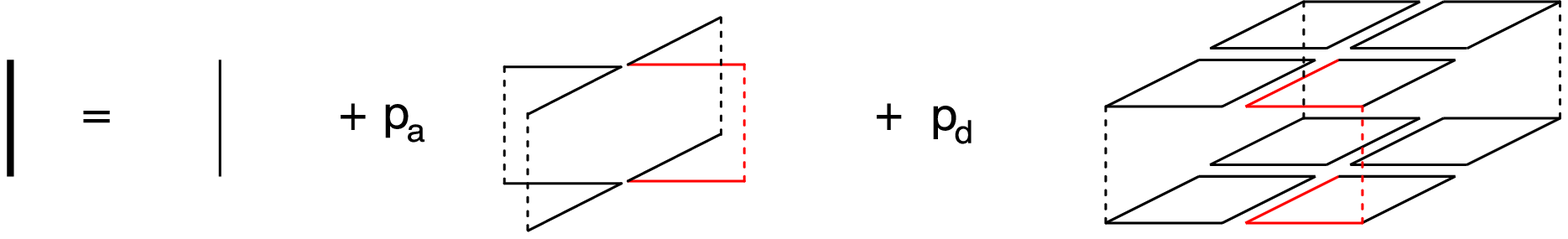,width=0.61\textwidth}
  \caption{Example of smearing of a single link. Sum of staples and
    diagonal staples are weighted with two independent parameters,
    respectively $p_a$ and $p_d$. }
  \label{fig:smear-path}
}
%\FIGURE[ht]{
%  \epsfig{figure=FIGS/blocking.eps,width=0.61\textwidth}
%  \caption{Blocking.}
%  \label{fig:block-path}
%}
\subsection*{Variational procedure}

Let us briefly describe the procedure we have used to obtain the mass of a lattice
state starting from a variational basis constructed as discussed
above. In the following, we denote by $\Phi_\alpha^{(R)}$ an operator
that creates a state in the $R$ channel, just as defined in
Eq.~(\ref{eq:linear-combination}), and the index $\alpha \in {1, \dotsc,N_o}$
labels one of the $N_o$ operators in the variational set. Since we
will be dealing with one channel at a time, from now on, we will drop the $(R)$
superscript for the sake of clarity.\\
The variational ansatz consists in finding an appropriate linear
combination of the basis operators 
\begin{equation}
  \label{eq:optimal-op}
  \tilde{\Phi}(t) \; = \; \sum_\alpha v_\alpha \Phi_\alpha(t)
  \quad ,
\end{equation}
whose correlator contains a single propagating state for which the
mass is minimised by the optimal $v_\alpha$. The optimisation of
$v_\alpha$ can be solved as a generalised eigenvalue problem.
We start by measuring a correlator matrix between all the possible
choices of operators in the variational set
\begin{equation}
  \label{eq:corr-matrix}
  \tilde{C}_{\alpha\beta}(t) \; = \; \sum_\tau \bra{0}
  \Phi_\alpha^\dag(t+\tau) \Phi_\beta(\tau) \ket{0}
  \quad .
\end{equation}
Then we diagonalise the matrix $\bar{C}(\bar{t})=\tilde{C}^{-1}(0) \tilde{C}(\bar{t})$
to solve the eigenvalue equation
\begin{equation}
  \label{eq:gen-eig-prob}
  \sum_\beta \bar{C}_{\alpha\beta}(\bar{t}) v_\beta \; = \; \sum_\beta
  e^{-m(\bar{t})\bar{t}}\bar{C}_{\alpha\beta} (0) v_\beta
  \quad ,
\end{equation}
where we choose $\bar{t}=1$. Each eigenvalue gives the mass of a
state in the chosen symmetry channel, thus the largest eigenvalue
corresponds to the ground state and its eigenvector is
used to construct the operator that best overlaps onto it following
Eq.~(\ref{eq:optimal-op}).
In the present work we studied the eigenvectors corresponding to the 5 highest
eigenvalues to construct operators $\tilde{\Phi}(t)_i$,
$i=1,\cdots,5$, whose diagonal correlators $C_{ii}(t)$ are fitted 
with the single-cosh ansatz
\begin{equation}
  \label{eq:cosh-ansatz}
C_{ii}(t) = |c_i|^2 \cosh \left( m_i t - N_L/2 \right)
\quad ,
\end{equation}
which assumes that the exponential of Eq~(\ref{eq:corr-easy}) is
dominated by only one state with mass $m_i$. This procedure allows us
to extract also masses of excited states since the operators
$\tilde{\Phi}_i(t)$ that decay with a mass $m_i \ne m_0$ within errors are generally good approximations of
excited glueballs wavefunctions.\\
A large variational set can result in a singular correlation matrix if
accidental linear dependencies between the operators develop. To
prevent singularities in the correlation matrix, we have defined
\emph{a posteriori} the variational basis in each channel as the
largest set of operators for which the inversion of $C$ can be
performed with a numerical accuracy of order $10^{-13}$ or better.

From the point of view of the implementation, it should be clear that
the simulation code must be able to handle a huge amount of different
shapes for the construction of the operators. Moreover, it is not
known \emph{a priori} which are the relevant shapes in the definition
of a good variational set. Thus the simulation should be flexible enough to allow us to change the used shapes with little effort.
In our project we have decided to develop a totally automated
environment that is able, at compile time, to create the code to
measure all the possible shapes in all the possible symmetry channels,
with the only input of a string representing each shape we want to
include in the calculation.
This environment has been created by means of a Mathematica code that,
starting from a set of shapes and symmetry channels, writes the plain functions to measure the properly symmetrized operators and their correlators. Such a solution turns out to be very efficient, since all the running time is used only in measuring the operators.
In the end our environment allows us to use a large basis of operators,
whose size is limited only by the amount of available memory and computer time, and that can hence be
increased at will with no programming effort.

\section{Numerical results}
\label{sect:numerical}
Our calculations are performed for $3 \le N \le 8$. As discussed in Sect.~\ref{sect:lattice}, at each $N$ we simulate at the coupling for which the system is at the deconfinement temperature on a $N_T = 6$ lattice. For $N=3,4,6,8$, $\beta_c(N_T = 6)$ is available from previous calculations~\cite{Lucini:2002ku,Lucini:2003zr,Datta:2009jn}. For the remaining values of $N$, we have computed $\beta_c(N_T)$ as follows~\cite{Lucini:2002ku,Lucini:2003zr}. On a lattice with geometry $N_L^3 \times N_T$ we computed first $\beta_c$ for $N_T = 5$ using an extrapolation to the thermodynamic limit of the position of the peak of the Polyakov loop susceptibility and then we assumed the coefficient of the $1/N_L^3$ term from that fit to extrapolate to the thermodynamic limit the peak position of the Polyakov loop susceptibility from results obtained at $N_T = 6$ at one value of $N_L$ in the asymptotic regime. Our analysis confirms the expectation that for $N=5$ and $N=7$ the deconfinement phase transition is first order; the details of this calculation will be reported elsewhere. Tab.~\ref{tab:summary-sim} summarises the relevant values of $\beta_c(N_T = 6)$ at the values of $N$ used in our investigation. The values of $\beta_c(N_T = 6)$ at $N=3,4,6,8$ are taken from Ref.~\cite{Lucini:2003zr} (Ref.~\cite{Datta:2009jn} quotes compatible results), while the values at $N=5,7$ have been computed as a part of this work.

The parameters used in the simulations are also summarised in Tab.~\ref{tab:summary-sim}. A typical run has $N_\tau$ thermalisation sweeps after which we start measuring the correlators. A Monte Carlo step consists of a compound sweep in which one heat--bath update is followed by $4$ over--relaxation sweeps. After the thermalisation process we perform $N_{MC}$ sweeps. We chose to measure every $N_{compound}$ sweeps to reduce autocorrelation between the measures. The $N_{measures}$ measure sweeps are further divided in $N_{bins}$ bins; each bin is an average over $N_{width}$ measures. The total set of measures to be analysed is then $N_{bins} \times \textrm{runs}$, where each run is independent of all the others. 

\TABLE{
\label{tab:summary-sim}
\begin{tabular}[c]{|c c c c c c c c c c|}
%number of colours ; beta ; lattice size ; measures ; sweeps ; compound
\hline
$N$ & $\beta_c(N_T = 6)$ & $\beta$ & $N_L$ & $N_\tau$ & $N_{MC}$ & $N_{compound}$ & $N_{width}$ & $N_{bins}$ & runs \\
\hline
3 & 5.8941(12)  & 5.8945 & 12 & 10k & 100k & 200 & 20 & 25 & 20 \\
4 & 10.7893(23) & 10.789 & 12 & 10k & 100k & 200 & 20 & 25 & 20 \\
5 & 17.1068(30) & 17.107   & 12 & 10k & 100k & 200 & 20 & 25 & 20 \\
6 & 24.8458(33) & 24.845 & 12 & 10k & 100k & 200 & 20 & 25 & 20 \\
7 & 33.9995(37) & 33.995 & 12 & 10k & 65k & 250 & 20 & 13 & 40 \\
8 & 44.4960(30) & 44.496   & 12 & 10k & 100k & 250 & 16 & 25 & 20 \\
\hline
\end{tabular}
\caption{Values of the critical couplings for the deconfinement temperature at $N_T = 6$ and parameters of the Monte Carlo simulations on lattices with $N_L^4$ points for $3 \le N \le 8$.}
}

After performing the variational calculation, the diagonal elements of the correlation matrix are fitted with the single-cosh ansatz of Eq.~(\ref{eq:cosh-ansatz}),
%\beq
%C(t) = |c_0|^2 \cosh \left( m_0 t - N_L/2 \right) \ ,
%\eeq
which assumes that only one state dominates the signal. Unitarity imposes $|c_i|^2 \le 1$. In practise, we are often able to obtain overlaps of the order of $0.95$, which proves the validity of the original variational ansatz. As a consequence, the fit generally works very well on the range $ 1 \le t \le 4$. 

In our investigation, we start by excluding at first the scattering states from our variational basis. Our numerical results are summarised in Tabs.~\ref{tab:su3}-\ref{tab:su8}, where we list for all the gauge groups the channels for which we were able to extract at least the groundstate mass; where more than a mass has been extracted, excitations are denoted with a number of $^\star$ indicating their position in the spectrum, according to the standard convention. For each gauge group, we are able to obtain a signal in most §of the $J^{PC}$ channels, with a good quality of the fit. We have highlighted in boldface the states for which  the $\chi^2$ is bigger than $1.25$; similarly, we have indicated in boldface states for which the overlap is less than $0.85$ (possibly indicating that the single-cosh fit might be a sub-optimal approximation) or statistically bigger than $1$ (the error on the overlap coming from the fit, which is not reported, is at most of the order of 0.1). Only the states that are free from those problems (which are a large majority) will be used for the large $N$ extrapolation.

The symmetry channel for which we are able to extract more states is the $A_1^{++}$. Symmetry channels whose groundstate is the same state in the continuum (e.g. the $E^{++}$ and the $T_2^{++}$, whose lowest-lying state is expected to be the $2^{++}$ glueball) have compatible groundstate mass with the expected degeneracy at our lattice spacing, confirming that for our choice of parameters we are in the scaling region.

The states obtained after the variational procedure can be decomposed into their projection onto the pure glueball states and onto the torelons:
\beq
\tilde{\Phi} = \alpha_G \Phi_G + \alpha_T \Phi_T \ ,
\eeq
where $\Phi_G$ and $\Phi_T$ are normalised to unity.
This allows us to define the pure glueball component and the torelon component of those states respectively as
\beq
\mbox{mix}_G = \frac{|\alpha_G|^2}{|\alpha_G|^2 + |\alpha_T|^2} \qquad \mbox{and} \qquad \mbox{mix}_T = \frac{|\alpha_T|^2}{|\alpha_G|^2 + |\alpha_T|^2} \ .
\eeq
For all gauge groups, there is a high mixing between narrow glueball trial states and torelon states in the first excitation of the $E^{++}$ and in the second excitation of the $A_1^{++}$. Other states with a consistent mixing with the torelons are the $T_2^{+-}$ and the $T_1^{--}$, the latter mostly for $N=3,4$. In order to guide the eye, in Tabs.~\ref{tab:su3}-\ref{tab:su8} we have labelled by (2T) the states that have an overlap with the torelons of order 0.35 or bigger and by (2T?) the states with an overlap between 0.15 and 0.35. For all other states, the overlap with the torelons is expected not to influence significantly the large-volume scaling of their mass.

We now turn to the problem of mixing of narrow resonances with scattering states. Since a calculation involving scattering states is much more demanding in terms of computer time, we use the results from the computation involving only single-particle and torelon operators to target the channels where mixing with multi-particle states is expected to affect significantly the results. At large $N$, this is expected to happen for the excited states that are close to twice the energy of the groundstate. It is then clear that the channel in which scattering states can potentially influence the measured spectrum in a relevant way is the $A_1^{++}$, where we can extract several excitations.
% Another interesting case is the $E_^{++}$ channel, in which finite-size effects show up in the form of a visible torelon; for this state, one can ask how it is going to be affected by mixing with scattering states.

In our investigation, we limit the study of scattering states to the $A_1$ channel, for which we perform a calculation on a larger set of operators, including multi-glueball operators in addition to the single-glueball and torelon operators used before. In fact, we perform calculations on separate sets of operators (the full set and the sets obtained excluding in turn scattering, torelon and single-glueball operators). Our results are reported in Tabs.~\ref{tab:a1pp-3-200}-\ref{tab:a1pp-7-200}. In analogy with the calculation involving only single-particle and torelon operators, we define the mixing coefficients mix$_G$, mix$_S$ and mix$_T$ that give the projection of the variational states respectively onto the single-glueball, scattering and torelon subspaces. The mass spectrum obtained with the different choices for the operator sets is illustrated in Figs.~\ref{fig:a1pp-3-200}-\ref{fig:a1pp-7-200}. The remarkable property shown by this calculation is that when only scattering and torelon operators are used the lowest-lying state has a mass that is much higher (roughly by a factor of two) than the mass of the groundstate extracted with the full variational basis. Moreover, the latter appears always when single-particle operators are included in the calculation. This is an indication that our multi-glueball set of operators project only on scattering states, as it should be. The scattering state seems to be slightly above the first excited single-glueball excitation at any value of $N$, and corresponds to the state identified as $A_1^{++\star\star}$ in the calculation that excludes the scattering operators (note the relatively large overlap of this state with the torelon operators in that calculation). Another feature we notice is that the number of excitations we are able to extract depends on the subset of operators used. This is hardly surprising, given the variational nature of the procedure. A similar analysis in the $A_1^{-+}$ channel shows that the $A_1^{-+\star}$ state identified with a variational calculation excluding scattering operators is in fact a scattering state.

We performed an additional check of finite volume effects in the $A_1^{++}$ channel studying SU(3) at $N_L = 18$, with the results illustrated in Fig.~\ref{fig:a1pp-3-18-200}. This test shows the expected result that
torelons are not visible anymore in the lowest-lying spectrum on this larger volume. In addition, the computed mass spectrum in the $A_1^{++}$ channel is compatible with the one measured for $N_L=12$. Since finite size effects are expected to be more relevant for the lightest glueballs, this check indicates that finite size corrections are negligible for $N_L = 12$.

In conclusion, our variational set enables us to correctly disentangle single-particle states from scattering states and torelon excitations. Moreover, the mixing coefficients give a measure of the contamination of the single-particle states from unwanted contributions. In the following, we consider single-glueball states all the states with a projection onto the single-glueball operator sector of 85\% or larger and we disregard states with a projection of 20\% or larger onto the torelon sector. Scattering states are identified as the states in the spectrum with a significant (30\% or larger) projection onto the scattering sector that are degenerate in mass with the lowest-lying state in the S + T basis when the latter has the largest component in the scattering sector. 

\section{The large $N$ spectrum}
\label{sect:lnspectrum}
In~\cite{Lucini:2001ej,Lucini:2004my} it has been shown that the lowest-lying states of the spectrum can be extrapolated to the large $N$ limit using only the leading large $N$ correction:
\beq
\label{eq:lnext}
a m_G (N)  = a m_G (\infty) + c/N^2 \ ,
\eeq
where $c$ is a coefficient of order one which depends on the symmetry channel. We perform a similar extrapolation for the states extracted in our work. We include in the fit all the states we have extracted from the variational procedure, indicating in boldface in the tables those who have been identified as potentially problematic in the previous section because of the overlap or of the larger $\chi^2$. The obtained large $N$ spectrum is reported in Tabs.~\ref{tab:large-a1}-\ref{tab:large-t2}\footnote{Note that for the $A_1^{++}$ channel we have indicated with $A_1^{++}$(S) the scattering states identified in the previous section and relabelled genuine single-particle excitations excluding these states from the spectrum. An analogous relabelling has been done in the $A_1^{-+}$ channel.}. Generally, when a sufficient number of reliable states is present, the inclusion of the more dubious states does not change the result within errors. This is shown explicitly in Tab.~\ref{tab:su-inf}, where the extrapolations with and without the less reliable states are compared. Both extrapolations are displayed in Figs.~\ref{fig:a1pp_ln}-\ref{fig:t2pm-mp-mm_ln} for all the quantum numbers for which a large $N$ limit can be extracted. We find that the ansatz~(\ref{eq:lnext}) works for all the measured states (including the excitations) for $N \ge 3$. In general, the central value of $c$ is found to be small (always of order one or below), as it is expected for a generic coefficient in a well-behaved expansion. For most of the states we find only modest corrections to the $N=\infty$ value of the mass: with a few exceptions, $c$ is compatible with zero and a fit with only the leading term $a m_G (\infty)$ in Eq.~(\ref{eq:lnext}) gives a result that is compatible with the fit that includes also the ${\cal O}(1/N^2)$ correction. In the following, we shall use the fit results obtained with the ansatz~(\ref{eq:lnext}).

As expected, the channel for which we extract the highest number of states is the $A_1^{++}$, where we clearly see the groundstate, the first two excitations and a scattering state. The large $N$ extrapolation is shown in Fig.~\ref{fig:a1pp_ln}. The groundstate is clearly visible also in the $A_1^{-+}$ channel, lying slightly below the first excitation of the $A_1^{++}$ channel (Fig.~\ref{fig:a1mp_ln}). Fig.~\ref{fig:a2pm_ln} shows the extrapolation of the groundstate in the $A_2^{+-}$ channel, which is the only state we can extract for the $A_2$ irreducible representation of the octahedral group. It is also clearly shown the effects of including less reliable masses in the extrapolation fit. In the $E$ channels, we get the groundstate and the first excitation for the $E^{++}$ (Fig.~\ref{fig:epp_ln}), in addition to a torelon state that we discard because it is not a spectral state. The ground state and the first excitation are also visible in the $E^{-+}$ (Fig.~\ref{fig:emp_ln}), while only the groundstate is present in our extrapolated data for the $E^{+-}$ and $E^{--}$ (Fig.~\ref{fig:epm-mm_ln}). The groundstate is clearly visible in the $T_1^{-+}$ and $T_1^{++}$ channels (Fig.~\ref{fig:t1pp-mp_ln}), while also the first excitation can be seen in the $T_1^{+-}$ channel (Fig.~\ref{fig:t1pm_ln}). Finally, two states appear in the $T_2^{++}$ channel (Fig.~\ref{fig:t2pp_ln}), with the groundstate determined for the $T_2^{-+}$, $T_2^{+-}$ and the $T_2^{--}$ channels (Fig.~\ref{fig:t2pm-mp-mm_ln}). The large $N$ extrapolation preserves the degeneracy of states that have the same spin in the continuum limit. The only exception to this general fact is observed for the ($E^{+-}$,$T_2^{+-}$) pair. We notice however that the extrapolation to $N \to \infty$ of the $T_2^{+-}$ groundstate only relies on two values at finite $N$. This probably explains the splitting between the two states at $N = \infty$.

We have also extrapolated to the large $N$ limit the scattering states identified in the $A_1^{++}$ and $A_1^{-+}$ channels. For the former state, for which we have better control, the extrapolated value of the energy is compatible with twice the mass of the groundstate, in agreement with the expectation that particles do not interact at large $N$. The energy of the $A_1^{-+}$(S) is slightly smaller than twice the mass of the corresponding groundstate, but this should probably be ascribed to the lack of control over the extrapolation to the large $N$ limit, which had to rely only on two finite-$N$ values.

The single-glueball spectrum determined in this work is plotted in Fig.~\ref{fig:spectrum_ln} and it is compared with the known spectrum at the same lattice spacing taken from Ref.~\cite{Lucini:2004my}. While the latter work achieve a better precision for the  $A_1^{++}$, the $E^{++}$ and the $A_1^{++\star}$, in this study we are able to measure seventeen more states. Moreover, our results for the common states are compatible with those reported in~\cite{Lucini:2004my}.

The determination of masses in various symmetry channels opens the possibility to study Regge trajectories at large $N$. A mandatory first step in this context is the identification of the continuum spin corresponding to each lattice state. The identification of the continuum spin is in general a very complicated task (see e.g. Ref.~\cite{Meyer:2004gx} for a discussion). In this work we use the simple-minded approach of looking at the subduced representations reported in Tab.~\ref{tab:subd-reps}, with the appearance of the same continuum state in various $\mathcal{G}_O$ representations taken into account whenever the expected degeneracies are unambiguous. We associate all states in the $A_1$ representation with spin $J=0$, the $T_1$ states are associated with $J=1$ and the only $J=3$ state is obtained from the $A_2$ lattice state. For $J=2$, we average the $E$ and the corresponding $T_2$ states when possible, except for $J^{PC} = 2^{+-}$ for which we take only the $E^{+-}$, having a poor control over the errors from the $T_2^{+-}$ extrapolation. The resulting Chew-Frautschi plot is shown in Fig.~\ref{fig:regge_new}. In order to obtain the string tension in lattice units at our lattice spacing, we have performed a large $N$ extrapolation of the data for $T_c/\sqrt{\sigma}$ in Refs.~\cite{Lucini:2003zr,Lucini:2005vg}, from which we obtain $T_c/\sqrt{\sigma} = 0.5851(32)$ for $N_T = 6$ and $N = \infty$. Although our results are in qualitative agreement with Ref.~\cite{Meyer:2004gx}, in order to check reliably predictions coming from various models more excitations and mostly higher spin states would be needed. 

\section{Conclusions}
\label{sect:conclusions}
In this work, we have studied numerically on the lattice the glueball spectrum in Yang-Mills SU($N$) gauge theories in the large $N$ limit. Using an automated technique for constructing trial wave functionals in all possible symmetry channels, we have built a large variational basis that has enabled us to obtain a large number of states, including some excitations. Moreover, the inclusion of functionals that best overlap with scattering and torelon states has allowed us to unambiguously exclude multi-particle states or finite-size artefacts from the spectrum of narrow resonances. This is a significant advance in our understanding of the large $N$ glueball spectrum from first principles. Our calculation being at fixed lattice spacing, a natural development of this work would be the determination of the spectrum in the continuum limit. We expect that as the lattice spacing becomes finer, more and more states will be pinned down by our variational calculation. In this respect, it might be helpful to use anisotropic lattices, like in Ref.~\cite{Morningstar:1997ff}. Together with a correct identification of the continuum spin states corresponding to the glueball states classified according to the irreducible representations of $\mathcal{G}_O$, a similar calculation would open the way to a comprehensive investigation of Regge trajectories, which are expected to be a more clean signature of the spectrum at large $N$ than at $N=3$. With little or no modification, the technique we have presented in this work will also prove helpful in related problems, like the lattice study of glueballs in QCD and the study of the low-energy spectrum of confining flux tubes.
\acknowledgments
We thank M. Peardon and M. Teper for discussions on the identification
of scattering states and on the construction of scattering
operators. Discussions with C. McNeile, H. Meyer, C. N\'unez and
A. Patella on various aspects of this work are also gratefully
acknowledged. Numerical simulations have been performed on a 120 core
Beowulf cluster partially funded by the Royal Society and STFC, and on
a 100 core cluster at Wuppertal University. The
work of B.L. is supported by the Royal Society through the University
Research Fellowship scheme and by STFC under contract ST/G000506/1. A.R.
thanks the Deutsche Forschungsgemeinschaft for financial support.
E.R. is supported by a SUPA Prize Studentship. E.R. acknowledges
financial support by the Royal Society in the early
stage of this work.

\bibliography{lnglueballs}

\newpage

%\section{Numerical results}
%\label{app:tables}
%~\\
%\hspace{-3cm}
\TABLE[bht]{
\label{tab:su3}
\begin{tabular}{|c|| c c c c c|}
% channel  ; mass(error) ; overlap ; chi square ; mixing g ;  mixing t
\hline
\multicolumn{6}{|c|}{SU(3) at $\beta=5.8945$ and $N_L=12$}\\
\hline
\hline
$R^{PC}$ &  $am_0$ & $|c_0|^2$ & $\chi^2$ & mix$_G$ &  mix$_T$ \\
\hline
$A_1^{++}$ &  0.798(15)  &  0.98  &  0.11  & 0.9932  &   0.0068 \\
$A_1^{++\star}$ & 1.336(45)  &  0.96  &  0.92  & 0.9542 &    0.0458 \\
$A_1^{++\star\star}$ & 1.674(79)  & 1.12 & 0.57 & 0.8563  &    0.1437 \\
$A_1^{++\star\star\star}$ & 1.85(13)  & 0.99 & 0.37  & 0.9667  &    0.0333 \\
$A_1^{++\star\star\star}$ & 1.96(18)  & 0.91 & 0.18  & 0.9430  &    0.0570 \\
$A_1^{-+}$ & 1.423(59)  &  0.90  &  0.63 & 0.9983  &    0.0017\\
\hline \hline
$A_2^{++}$(2T) & 1.81(12)  &  0.90  &  0.01 & 0.0509  &   0.9491 \\
$A_2^{++}$(2T) & 2.70(68)  &  0.92  &  0.18 & 0.2069  &   0.7931\\
$A_2^{+-}$ & 2.33(38)& 0.85 & 0.43 & 0.9982 &0.0018\\
$A_2^{--}$ & 2.41(42) & 0.85 & 0.01 & 0.8855 &0.1145\\
\hline \hline
$E^{++}$ & 1.253(35) & 0.96  & \bf{1.67} & 0.9224  &     0.0776\\
$E^{++}$ & 1.290(39) & 0.97 & 0.04 & 0.9098  &    0.0902\\
$E^{++}$(2T) & 1.374(64)  & 0.91 & \bf{1.42} & 0.1649  &     0.8351\\
$E^{++}$(2T) & 1.554(62) & 1.05 & \bf{1.39} & 0.3735  &     0.6265\\
$E^{+-}$ & 2.57(44) & 1.01  &  0.90 &  0.9818 &      0.0182\\
$E^{--}$ & 2.22(27) & 0.98  &  1.24 & 0.9956 &      0.0044\\
$E^{--}$ & 2.15(22) & 0.90  &  0.35 & 0.9528 &      0.0472\\
$E^{-+}$ & 1.656(89) &  0.92  &  0.39 & 0.9982  &    0.0018\\
$E^{-+}$ & 1.623(83)  & 0.87  &  0.66 & 0.9962 &     0.0038\\
$E^{-+\star}$ & 2.37(33)  & 1.01  &  0.54 & 0.9738  &  0.0074\\
$E^{-+\star}$ & 2.43(38) & 1.01 & 0.51 & 0.9994 &  0.0006 \\
$E^{-+\star}$ & 2.53(57) & 0.89 & 0.26  & 0.9551 &  0.0449 \\
\hline \hline
$T_1^{++}$ & 1.97(17) & 0.86 & 0.09 & 0.9964 &  0.0036 \\
$T_1^{++}$ & 2.05(21) & 0.87 & 0.11  & 0.9871 &  0.0129 \\
$T_1^{+-}$ & 1.611(78) & 0.97 & 0.30 & 0.9777 &  0.0223 \\
$T_1^{+-}$ & 1.563(79) & 0.89 & \bf{1.30} & 0.9245 &  0.0755 \\
$T_1^{+-}$(2T) & 1.92(15) & 0.94 & \bf{1.61} & 0.5869 &  0.4131 \\
$T_1^{--}$(2T) & 2.36(30) & 1.06 & 0.16 & 0.5170  &  0.4830 \\
$T_1^{--}$(2T) & 2.35(33) & 0.99 & 0.43 & 0.6290  &  0.3710 \\
$T_1^{--}$(2T) & 2.42(37) & 1.06 & 0.35 & 0.6400  &  0.3600 \\
\hline \hline
$T_2^{++}$ & 1.292(41) & 0.98 & 0.41 & 0.9746 &  0.0254 \\
$T_2^{++}$ & 1.324(42) & 0.98 & 0.61 & 0.9828 &  0.0172 \\
$T_2^{++}$ & 1.79(11)  &  0.97&  0.16 & 0.9635 &  0.0365 \\
$T_2^{++}$ & 1.79(11)  &  0.96&  0.38 & 0.9887 &  0.0113 \\
$T_2^{+-}$(2T) & 1.87(14)  &  0.91  &  0.33  & 0.5195  &    0.4805 \\
$T_2^{+-}$(2T?) & 1.85(13)  &  0.89  &  1.00  & 0.7641  &    0.2359 \\
$T_2^{+-}$(2T?) & 2.13(19)  &  1.12  &  0.19  & 0.6719  &    0.3281 \\
$T_2^{--}$  & 2.26(28)  & 0.97 & 0.18 & 0.9968  &    0.0032 \\
$T_2^{-+}$ & 1.665(93)  &  0.94  &  0.73 &  0.9803  &    0.0197 \\
$T_2^{-+}$ & 1.614(82)  &  0.89  &  0.01 &  0.9788  &    0.0212 \\
$T_2^{-+}$ & 2.13(26) &  0.88  &  0.70 &  0.9722  &    0.0278 \\
\hline
\end{tabular}
\caption{The measured spectrum of pure SU(3) gauge theory from lattice
  simulations at the parameters shown in the header.}
}

\TABLE[ht]{
\label{tab:su4}
\begin{tabular}[c]{|c|| c c c c c|}
% channel ; #eigenv ; mass ; error ; overlap ; chi square ; mixing g ;  mixing t
\hline
\multicolumn{6}{|c|}{SU(4) at $\beta=10.789$ and $N_L=12$}\\
\hline
\hline
$R^{PC}$ & $am_0$ & $|c_0|^2$ & $\chi^2$ &mix$_G$ &  mix$_T$ \\
\hline
$A_1^{++}$ & 0.821(15)  &  0.99  &  1.03 &  0.9928 &    0.0072 \\
$A_1^{++\star}$ & 1.415(47)  &  0.96  &  \bf{1.68} &  0.9871 &    0.0129 \\
$A_1^{++\star}$(2T?) & 1.596(73)  &  0.97  &  0.74 &  0.7871 &    0.2129 \\
$A_1^{++\star\star}$ & 1.97(18)  &  0.85  &  0.43 &  0.9842 &    0.0158 \\
$A_1^{+-}$ & 2.79(86)  &  0.86  &  0.04 &  1.0000 &    0.0000 \\
$A_1^{-+}$ & 1.423(54)  &  0.89  &  0.21 &  0.9991 &    0.0009 \\
$A_1^{-+\star}$ & 2.67(63)  &  1.08  &  0.67 &  0.9927 &    0.0073 \\
\hline \hline
$A_2^{++}$(2T) & 1.83(15)  &  0.85  &  \bf{1.55} &  0.0271 &    0.9729 \\
$A_2^{++\star}$(2T?) & 2.66(66)  &  0.87  &  0.11 &  0.8390 &    0.1610 \\
$A_2^{+-}$ & 2.08(19)  &  0.99  &  0.27 &  0.9998 &    0.0002 \\
$A_2^{--}$ & 2.34(37)  &  0.87  &  0.47 &  0.9291 &    0.0709 \\
\hline \hline
$E^{++}$ & 1.270(41)  &  0.94  &  0.01 &  0.9787 &    0.0213 \\
$E^{++}$ & 1.346(47)  &  1.00  &  0.46 &  0.9553 &    0.0447 \\
$E^{++}$(2T) & 1.634(75)  &  1.03  &  0.43 &  0.3336 &    0.6664 \\
$E^{++}$(2T) & 1.711(91)  &  1.07  &  0.01 &  0.2347 &    0.7653 \\
$E^{+-}$ & 2.85(70)  &  \bf{1.17}  &  0.66 &  0.9728 &    0.0272 \\
$E^{-+}$ & 1.732(97)  &  0.95  &  0.03 &  0.9922 &    0.0078 \\
$E^{-+}$ & 1.653(92)  &  \bf{0.84}  &  0.48 &  0.9953 &    0.0047 \\
\hline \hline
$T_1^{++}$ & 2.13(22)  &  0.88  &  0.35 &  0.9911 &    0.0089 \\
$T_1^{+-}$ & 1.638(83)  &  0.99  &  1.19 &  0.9867 &    0.0133 \\
$T_1^{+-}$ & 1.639(77)  &  0.98  &  0.20 &  0.9634 &    0.0366 \\
$T_1^{+-}$ & 1.718(88)  &  1.03  &  0.01 &  0.9657 &    0.0343 \\
$T_1^{+-}$ & 1.79(13)  &  0.85  &  0.05 &  0.9464 &    0.0536 \\
$T_1^{+-\star}$ & 2.15(19)  &  1.15  &  0.02 &  0.9522 &    0.0478 \\
$T_1^{--}$(2T) & 2.17(28)  &  0.88  &  0.45 &  0.5640 &    0.4360 \\
$T_1^{--}$(2T) & 2.45(39)  &  0.99  &  0.02 &  0.4546 &    0.5454 \\
$T_1^{-+}$ & 2.37(36)  &  0.90  &  0.02 &  0.9842 &    0.0158 \\
\hline \hline
$T_2^{++}$ & 1.333(47)  &  0.99  &  0.01 &  0.9858 &    0.0142 \\
$T_2^{++\star}$ & 1.76(11)  &  0.89  &  0.17 &  0.9910 &    0.0090 \\
$T_2^{+-}$(2T) & 1.96(16)  &  0.90  &  0.48 &  0.6320 &    0.3680 \\
$T_2^{+-}$(2T) & 2.24(23)  &  \bf{1.16}  &  0.20 &  0.5183 &    0.4817 \\
$T_2^{+-}$(2T) & 2.12(20)  &  1.00  &  0.22 &  0.2399 &    0.7601 \\
$T_2^{--}$ & 2.21(25)  &  0.89  &  \bf{1.75} &  0.9883 &    0.0117 \\
$T_2^{-+}$ & 1.609(87)  &  0.86  &  0.04 &  0.9842 &    0.0158 \\
$T_2^{-+}$ & 1.678(95)  &  0.88  &  0.01 &  0.9597 &    0.0403 \\
\hline
\end{tabular}
\caption{The measured spectrum of pure SU(4) gauge theory from lattice
  simulations at the parameters shown in the header.}
}

\TABLE[ht]{
\label{tab:su5}
\begin{tabular}[c]{|c|| c c c c c|}
% channel ; #eigenv ; mass ; error ; overlap ; chi square ; mixing g ;  mixing t
\hline
\multicolumn{6}{|c|}{SU(5) at $\beta=17.107$ and $N_L=12$}\\
\hline
\hline
$R^{PC}$ & $am_0$ & $|c_0|^2$ & $\chi^2$ & mix$_G$ &  mix$_T$ \\
\hline
$A_1^{++}$ & 0.798(14) &  0.98 &  0.70 & 0.9901 &   0.0099 \\
$A_1^{++\star}$ & 1.426(57) &  0.92 &  0.01 & 0.9864 &   0.0136 \\
$A_1^{++}$(2T?) & 1.675(99) &  0.94 &  \bf{1.61} & 0.7006 &   0.2994 \\
$A_1^{++\star\star}$ & 2.03(20) &  0.88&  0.40 & 0.9986 &   0.0014 \\
$A_1^{-+}$ & 1.406(57) &  0.88 &  0.02 & 0.9975 &   0.0025 \\
$A_1^{-+\star}$ & 2.28(29) &  0.92 &  0.48 & 0.9877 &   0.0123 \\
\hline \hline
$A_2^{++}$(2T) & 1.88(16) &  0.85 &  0.08 & 0.0750 &   0.9250 \\
$A_2^{+-}$ & 2.19(22) &  1.11 &  0.35 & 0.9980 &   0.0020 \\
\hline \hline
$E^{++}$ & 1.300(42) &  0.95 &  1.00 & 0.9830 &   0.0170 \\
$E^{++}$(2T) & 1.628(87) &  0.94 &  \bf{1.73} & 0.2965 &   0.7035 \\
$E^{++}$(2T) & 1.597(84) &  0.89 &  0.03 & 0.5388 &   0.4612 \\
$E^{++\star}$ & 1.91(14) &  1.05 &  0.20 & 0.9619 &   0.0381 \\
$E^{--}$ & 2.05(22) &  \bf{0.83} &  0.01 & 0.9737 &   0.0263 \\
$E^{-+}$ & 1.72(10) &  0.90 &  0.01 & 0.9905 &   0.0095 \\
$E^{-+\star}$ & 2.48(43) &  0.97 &  0.05 & 0.9956 &   0.0044 \\
\hline \hline
$T_1^{++}$ & 2.09(21) &  0.91 &  0.03 & 0.9952 &   0.0048 \\
$T_1^{++}$ & 2.03(21) &  \bf{0.82} &  0.30 & 0.9961 &   0.0039 \\
$T_1^{+-}$ & 1.598(80) &  0.95 &  0.00 & 0.9820 &   0.0180 \\
$T_1^{+-}$ & 1.625(76) &  0.96 &  0.82 & 0.9570 &   0.0430 \\
$T_1^{+-\star}$ & 1.99(16) &  0.99 &  0.11 & 0.9107 &   0.0893 \\
$T_1^{--}$(2T?) & 2.16(28) &  \bf{0.81} &  0.70 & 0.8117 &   0.1883 \\
$T_1^{-+}$ & 2.30(30) &  0.90 &  1.19 & 0.9953 &   0.0047 \\
\hline \hline
$T_2^{++}$ & 1.261(41) &  0.90 &  0.13 & 0.9738 &   0.0262 \\
$T_2^{++}$ & 1.321(45) &  0.94 &  0.03 & 0.9784 &   0.0216 \\
$T_2^{++}$ & 1.391(48) &  1.02 &  0.19 & 0.9931 &   0.0069 \\
$T_2^{++\star}$ & 1.76(12) &  0.86 &  0.39 & 0.9809 &   0.0191 \\
$T_2^{++\star}$ & 1.91(14) &  0.99 &  0.77 & 0.9798 &   0.0202 \\
$T_2^{+-}$ & 1.90(15) &  \bf{0.84} &  0.55 & 0.9670 &   0.0330 \\
$T_2^{+-}$(2T?) & 1.94(15) &  0.94 &  0.03 & 0.8434 &   0.1566 \\
$T_2^{+-}$(2T) & 2.03(20) &  0.87 &  0.18 & 0.2209 &   0.7791 \\
$T_2^{--}$ & 2.01(21) &  \bf{0.80} &  0.65 & 0.9955 &   0.0045 \\
$T_2^{-+}$ & 1.71(10) &  0.92 &  0.71 & 0.9865 &   0.0135 \\
$T_2^{-+\star}$ & 1.93(12) &  1.13 &  0.56 & 0.9895 &   0.0105 \\
$T_2^{-+\star}$ & 2.31(31) &  0.95 &  0.08 & 0.9601 &   0.0399 \\
$T_2^{-+\star}$ & 2.29(29) &  0.91 &  0.86 & 0.9490 &   0.0510 \\
\hline
\end{tabular}
\caption{The measured spectrum of pure SU(5) gauge theory from lattice
  simulations at the parameters shown in the header.}
}

\TABLE[ht]{
\label{tab:su6}
\begin{tabular}[c]{|c|| c c c c c|}
% channel ; #eigenv ; mass ; error ; overlap ; chi square ; mixing g ;  mixing t
\hline
\multicolumn{6}{|c|}{SU(6) at $\beta=24.845$ and $N_L=12$}\\
\hline
\hline
$R^{PC}$ & $am_0$ & $|c_0|^2$ & $\chi^2$ & mix$_G$ &  mix$_T$ \\
\hline
$A_1^{++}$ & 0.785(14)  & 0.96  & 0.20 & 0.9959  &  0.0041 \\
$A_1^{++}$(2T) & 1.572(83)  & \bf{0.82}  & 0.54 & 0.5763  &  0.4237 \\
$A_1^{++\star}$ & 2.37(28)  & \bf{1.19}  & 0.05 & 0.9842  &  0.0158 \\
$A_1^{-+}$ & 1.417(59)  & 0.89  & \bf{1.51} & 0.9996  &  0.0004 \\
\hline \hline
$A_2^{+-}$ & 1.85(14)  & \bf{0.80}  & 0.74 & 0.9993  &  0.0007 \\
$A_2^{+-}$ & 2.59(51)  & 0.99  & 0.03 & 0.9991  &  0.0009 \\
\hline \hline
$E^{++}$ & 1.257(43)  & 0.89  & 0.40 & 0.9728  &  0.0272 \\
$E^{++}$(2T) & 1.607(86)  & 0.92  & 0.83 & 0.2433  &  0.7567 \\
$E^{++\star}$(2T?) & 1.97(14)  & 1.09  & 0.16 & 0.8404  &  0.1596 \\
$E^{--}$ & 2.33(28)  & 1.07  & 0.02 & 0.9913  &  0.0087 \\
$E^{-+}$ & 1.76(11)  & 0.93  & 0.75 & 0.9929  &  0.0071 \\
$E^{-+}$ & 1.71(11)  & 0.85  & 1.16 & 0.9951  &  0.0049 \\
$E^{-+\star}$ & 2.56(45)  & 1.07  & 0.01 & 0.9886  &  0.0114 \\
\hline \hline
$T_1^{++}$ & 2.30(28)  & 0.99  & 0.01 & 0.9924  &  0.0076 \\
$T_1^{++}$ & 2.29(33)  & 0.92  & 0.66 & 0.9919  &  0.0081 \\
$T_1^{+-}$ & 1.501(66)  & 0.86  & 0.45 & 0.9860  &  0.0140 \\
$T_1^{+-}$ & 1.658(77) & 0.97  & 0.66 & 0.9539  &  0.0461 \\
$T_1^{+-\star}$ & 1.94(16)  & 0.93  & 0.11 & 0.9443  &  0.0557 \\
$T_1^{-+}$ & 2.43(43)  & 0.93  & 0.03 & 0.9859  &  0.0141 \\
\hline \hline
$T_2^{++}$ & 1.263(39)  & 0.90  & 0.74 & 0.9881  &  0.0119 \\
$T_2^{++}$ & 1.367(46)  & 0.99  & \bf{1.96} & 0.9841  &  0.0159 \\
$T_2^{++\star}$ & 1.99(14)  & 1.05  & 0.56 & 0.9704  &  0.0296 \\
$T_2^{++\star}$ & 2.02(16)  & 1.07  & \bf{1.71} & 0.9879  &  0.0121 \\
$T_2^{+-}$ & 1.98(16)  & 0.93  & 0.14 & 0.8618  &  0.1382 \\
$T_2^{+-}$ & 2.08(18)  & 1.02  & 0.78 & 0.9465  &  0.0535 \\
$T_2^{--}$ & 2.31(30)  & 0.99  & 0.15 & 0.9972  &  0.0028 \\
$T_2^{-+}$ & 1.85(12)  & 1.05  & 0.18 & 0.9416  &  0.0584 \\
$T_2^{-+}$ & 1.74(11)  & 0.91  & 0.13 & 0.9765  &  0.0235 \\
$T_2^{-+}$ & 1.63(10)  & \bf{0.80}  & 0.88 & 0.9790  &  0.0210 \\
\hline
\end{tabular}
\caption{The measured spectrum of pure SU(6) gauge theory from lattice
  simulations at the parameters shown in the header.}
}

\TABLE[ht]{
\label{tab:su7}
\begin{tabular}[c]{|c|| c c c c c|}
% channel ; #eigenv ; mass ; error ; overlap ; chi square ; mixing g ;  mixing t
\hline
\multicolumn{6}{|c|}{SU(7) at $\beta=33.995$ and $N_L=12$}\\
\hline
\hline
$R^{PC}$ & $am_0$ & $|c_0|^2$ & $\chi^2$ & mix$_G$ & mix$_T$ \\
\hline
$A_1^{++}$ & 0.820(15) &  0.98 &  0.43 & 0.9976 &  0.0024 \\
$A_1^{++\star}$ & 1.564(61) &  1.08 &  1.08 & 0.9920 &  0.0080 \\
$A_1^{++}$(2T) & 1.744(89) &  0.99 &  \bf{1.29} & 0.6089 &  0.3911 \\
$A_1^{++\star\star}$ & 2.30(27) &  1.06 &  0.07 & 0.9970 &  0.0030 \\
$A_1^{-+}$ & 1.452(58) &  0.92 &  0.07 & 0.9994 &  0.0006 \\
$A_1^{-+\star}$ & 2.28(32) &  0.89 &  0.23 & 0.9968 &  0.0032 \\
\hline \hline
$A_2^{++}$(2T) & 1.84(13) &  \bf{0.78} &  1.12 & 0.0416 &  0.9584 \\
$A_2^{++}$ & 2.18(25) &  0.88 &  0.44 & 0.9231 &  0.0769 \\
$A_2^{+-}$ & 2.43(43) &  \bf{0.84} &  0.23 & 0.9996 &  0.0004 \\
\hline \hline
$E^{++}$ & 1.296(44) &  0.95 &  0.03 & 0.9841 &  0.0159 \\
$E^{++}$ & 1.303(46) &  0.93 &  \bf{1.31} & 0.9723 &  0.0277 \\
$E^{++}$(2T) & 1.649(89) &  0.92 &  \bf{1.44} & 0.2811 &  0.7189 \\
$E^{++}$(2T) & 1.72(10) &  0.93 &  1.19 & 0.3631 &  0.6369 \\
$E^{++\star}$ & 1.81(12) &  0.91 &  0.73 & 0.8825 &  0.1175 \\
$E^{+-}$ & 2.58(62) &  0.87 &  0.39 & 0.9948 &  0.0052 \\
$E^{+-}$ & 2.63(64) &  0.85 &  0.04 & 0.9950 &  0.0050 \\
$E^{--}$ & 2.42(30) &  1.11 &  \bf{1.50} & 0.9904 &  0.0096 \\
$E^{--}$(2T) & 2.79(66) &  1.05 &  0.23 & 0.5190 &  0.4810 \\
$E^{-+}$ & 1.78(11) &  0.91 &  0.24 & 0.9974 &  0.0026 \\
$E^{-+\star}$ & 2.59(42) &  1.09 &  0.66 & 0.9958 &  0.0042 \\
\hline \hline
$T_1^{++}$ & 2.27(25) &  0.99 &  \bf{1.92} & 0.9962 &  0.0038 \\
$T_1^{++}$ & 2.42(31) &  1.15 &  0.44 & 0.9874 &  0.0126 \\
$T_1^{+-}$ & 1.543(73) &  0.89 &  0.11 & 0.9628 &  0.0372 \\
$T_1^{+-}$ & 1.617(82) &  0.93 &  0.52 & 0.9581 &  0.0419 \\
$T_1^{+-}$ & 1.596(79) &  0.90 &  0.15 & 0.9734 &  0.0266 \\
$T_1^{+-\star}$ & 1.89(15) & 0.85 &  0.01 & 0.9895 &  0.0105 \\
$T_1^{--}$ & 2.12(22) &  0.87 &  0.60 & 0.9334 &  0.0666 \\
$T_1^{--}$ & 2.46(33) &  1.11 &  0.88 & 0.9548 &  0.0452 \\
$T_1^{--}$ & 2.33(37) &  \bf{0.84} &  0.07 & 0.9092 &  0.0908 \\
$T_1^{-+}$ & 2.48(46) &  0.99 &  \bf{1.98} &  0.9605 &  0.0395 \\
\hline \hline
$T_2^{++}$ & 1.322(45) &  0.95 &  0.24 & 0.9867 &  0.0133 \\
$T_2^{++}$ & 1.367(50) &  0.97 &  0.99 & 0.9879 &  0.0121 \\
$T_2^{++}$ & 1.437(50) &  1.04 &  \bf{1.73} & 0.9799 &  0.0201 \\
$T_2^{++\star}$ & 1.96(15) &  0.99 &  0.21 & 0.9780 &  0.0220 \\
$T_2^{+-}$ & 2.07(17) &  0.98 &  \bf{1.27} & 0.9682 &  0.0318 \\
$T_2^{+-}$(2T) & 2.05(18) &  0.93 &  \bf{1.73} & 0.6236 &  0.3764 \\
$T_2^{+-}$(2T) & 2.23(23) &  1.04 &  0.21 & 0.2859 &  0.7141 \\
$T_2^{--}$ & 2.45(33) &  1.13 &  0.20 & 0.9886 &  0.0114 \\
$T_2^{-+}$ & 1.77(11) &  0.97 &  0.40 & 0.9738 &  0.0262 \\
$T_2^{-+\star}$ & 2.36(34) &  0.98 &  1.11 & 0.9618 &  0.0382 \\
\hline
\end{tabular}
\caption{The measured spectrum of pure SU(7) gauge theory from lattice
  simulations at the parameters shown in the header.}
}

\TABLE[ht]{
\label{tab:su8}
\begin{tabular}[c]{|c|| c c c c c|}
% channel ; #eigenv ; mass ; error ; overlap ; chi square ; mixing g ;  mixing t
\hline
\multicolumn{6}{|c|}{SU(8) at $\beta=44.496$ and $N_L=12$}\\
\hline
\hline
$R^{PC}$ & $am_0$ & $|c_0|^2$ & $\chi^2$ & mix$_G$ &  mix$_T$ \\
\hline
$A_1^{++}$ & 0.785(16) &  0.96 &  0.77 & 0.9993 &  0.0007 \\
$A_1^{++\star}$ & 1.408(63) &  0.90 &  0.52 & 0.9966 &  0.0034 \\
$A_1^{-+}$ & 1.401(62) &  \bf{0.83} &  \bf{1.48} & 0.9995 &  0.0005 \\
\hline \hline
$A_2^{+-}$ & 2.21(27) &  1.08 &  \bf{2.38} & 0.9981 &  0.0019 \\
\hline \hline
$E^{++}$ & 1.339(53) &  0.96 &  0.38 & 0.9937 &  0.0063 \\
$E^{++}$(2T) & 1.78(12) &  0.94 &  0.34 & 0.5444 &  0.4556 \\
$E^{++\star}$ & 1.80(14) &  0.88 &  0.44 & 0.9330 &  0.0670 \\
$E^{+-}$ & 2.67(72) &  0.93 &  0.02 & 0.9815 &  0.0185 \\
$E^{--}$ & 2.24(30)  & 0.92 &  \bf{1.59} & 0.9876 &  0.0124 \\
$E^{--}$ & 2.50(41)  & 1.15 &  0.27 & 0.9806 &  0.0194 \\
$E^{-+}$ & 1.90(14) &  1.05 &  0.55 & 0.9976 &  0.0024 \\
\hline \hline
$T_1^{++}$ & 2.25(33) &  0.87 &  0.32 & 0.9915 &  0.0085 \\
$T_1^{+-}$ & 1.693(97) &  1.00 &  0.07 & 0.9742 &  0.0258 \\
$T_1^{+-}$ & 1.682(97) &  0.98 &  \bf{1.28} & 0.9847 &  0.0153 \\
$T_1^{+-}$ & 1.590(99) &  0.88 &  0.09 & 0.9354 &  0.0646 \\
$T_1^{+-\star}$ & 1.98(17) &  0.97 &  0.60 & 0.9748 &  0.0252 \\
$T_1^{+-\star}$ & 2.10(24) &  1.04 &  0.52 & 0.9803 &  0.0197 \\
$T_1^{--}$(2T?) & 2.16(27) &  0.88&  0.70 & 0.7766 &  0.2234 \\
$T_1^{--}$(2T?) & 2.47(43) &  1.14 &  1.15 & 0.8164 &  0.1836 \\
$T_1^{-+}$ & 2.45(43) &  0.95 &  0.80 & 0.9909 &  0.0091 \\
$T_1^{-+}$ & 2.57(57) &  0.98 &  0.53 & 0.9931 &  0.0069 \\
\hline \hline
$T_2^{++}$ & 1.323(51) &  0.93 &  0.00 & 0.9853 &  0.0147 \\
$T_2^{++}$ & 1.451(56) &  1.05 &  1.23 & 0.9687 &  0.0313 \\
$T_2^{++}$ & 1.393(57) &  0.97 &  0.39 & 0.9749 &  0.0251 \\
$T_2^{++\star}$ & 1.90(15) &  0.93 &  \bf{1.67} & 0.9878 &  0.0122 \\
$T_2^{+-}$ & 2.15(21) &  1.12 &  0.72 & 0.9232 &  0.0768 \\
$T_2^{+-}$ & 2.07(21) &  0.97 &  0.02 & 0.9754 &  0.0246 \\
$T_2^{+-}$ & 1.99(18) &  0.89 &  1.25 & 0.9644 &  0.0356 \\
$T_2^{+-}$(2T) & 2.11(24) &  0.92 &  \bf{1.41} & 0.3123 &  0.6877 \\
$T_2^{+-}$(2T) & 2.26(28) &  1.01 &  1.25 & 0.3711 &  0.6289 \\
$T_2^{--}$ & 2.20(32) &  0.93 &  0.54 & 0.9870 &  0.0130 \\
$T_2^{--}$ & 2.12(31) &  \bf{0.82} &  1.05 & 0.9904 &  0.0096 \\
$T_2^{-+}$ & 2.01(16) &  \bf{1.19} &  0.02 & 0.9786 &  0.0214 \\
$T_2^{-+}$ & 1.79(14) &  0.95 &  0.95 & 0.9628 &  0.0372 \\
$T_2^{-+}$ & 2.22(33) &  0.87 &  0.41 & 0.9306 &  0.0694 \\
\hline
\end{tabular}
\caption{The measured spectrum of pure SU(8) gauge theory from lattice
  simulations at the parameters shown in the header.}
}
\clearpage
\TABLE{
\label{tab:a1pp-3-200}
\begin{tabular}[c]{|c|| c c c c c c |}
\hline
\multicolumn{7}{|c|}{SU($3$) at $\beta=5.8945$ and $N_L=12$}\\
\hline
\hline
$R^{PC}$ &  $am(\sigma)$ & $|c_n|^2$ & $\chi^2$ & mix$_G$ & mix$_S$ & mix$_T$ \\
\hline
\hline
$A_1^{++}$                & 0.792(14)  &0.98  &0.02 &0.9664  &  0.0180  &  0.0156\\
$A_1^{++\star}$           & 1.370(50)  &  1.01  &   0.01 &0.8242  &  0.0870  &  0.0888\\
$A_1^{++\star\star}$      & 1.609(97)  &  0.90  &   0.74 &0.8060  &  \bf{0.1753}  &  0.0187\\
$A_1^{++\star\star\star}$ & 1.86(13)  &  0.99  &   0.05 &0.9197  &  0.0531  &  0.0273\\
\hline
\hline
$A_1^{++}$                &  0.792(14)  &  0.98  &   0.02 &0.9830   & 0.0170    &--\\
$A_1^{++\star}$           &  1.385(51)  &  1.01  &   0.17 &0.9129   & 0.0871    &--\\
$A_1^{++\star\star}$      &  1.630(97)  &  0.89  &   0.07 &0.4183  &  \bf{0.5817}    &--\\
$A_1^{++\star\star\star}$ &1.755(98)  &  1.07  &   0.23 &0.9563 &   0.0437    &--\\
$A_1^{++\star\star\star\star}$ & 1.95(16)  &  0.97  &   0.38 &0.9647   & 0.0353    &--\\
\hline
\hline
$A_1^{++}$                & 0.792(14)  &  0.98  &   0.03 &0.9842    &--&    0.0158\\
$A_1^{++\star}$           & 1.405(51)  &  1.02  &   0.09 &0.9249    &--&    0.0751\\
$A_1^{++\star\star}$      & 1.93(14)  &  1.08  &   0.06 &0.9718    &--&    0.0282\\
\hline
\hline
$A_1^{++}$                & 1.485(82)  &  \bf{0.82}  &   0.06 &--&    0.8321   & 0.1679\\
$A_1^{++\star}$           &1.94(19)  &  0.99  &   0.47 &--&    0.2613  &  0.7387\\
\hline
\hline
\end{tabular}
\caption{Masses in the $A_1^{++}$ channel at the parameters shown in the header. A wide range of different
  variational sets is employed and the results are compared. In
  particular, we want to highlight the mixing with multi--glueballs
  and bi--torelon states.}
}
\TABLE{
\label{tab:a1pp-3-18-200}
\begin{tabular}[c]{|c|| c c c c c c |}
\hline
\multicolumn{7}{|c|}{SU($3$) at $\beta=5.8945$ and $N_L=18$}\\
\hline
\hline
$R^{PC}$ &  $am(\sigma)$ & $|c_n|^2$ & $\chi^2$ & mix$_G$ & mix$_S$ & mix$_T$ \\
\hline
\hline
$A_1^{++}$                & 0.799(13)  & 0.98  & 0.54 & 0.9618 & 0.0107 & 0.0276\\
$A_1^{++\star}$           & 1.350(39)  & 0.92  & 0.57 & 0.9819 & 0.0133 & 0.0048\\
$A_1^{++\star\star}$      & 1.546(71)  & 0.87  & 1.14 & 0.6752 & \bf{0.3098} & 0.0150\\
$A_1^{++\star\star\star}$ & 1.90(13)  & \bf{0.84}  & 0.83 & 0.9917 & 0.0065 & 0.0019\\
\hline
\hline
$A_1^{++}$                & 0.799(13)  & 0.98  & 0.62 & 0.9904 & 0.0096& -- \\
$A_1^{++\star}$           & 1.347(39) & 0.91  & 0.67 & 0.9873 & 0.0127& -- \\
$A_1^{++\star\star}$      & 1.543(70) & 0.87  & 1.10 & 0.6851 & \bf{0.3149}& -- \\
$A_1^{++\star\star\star}$ & 1.91(13) & \bf{0.84}  & 0.83 & 0.9936 & 0.0064& -- \\
\hline
\hline
$A_1^{++}$                & 0.799(13)  & 0.98  & 0.52 & 0.9801 & -- & 0.0199\\
$A_1^{++\star}$           & 1.353(39)  & 0.91  & 0.60 & 0.9967 & -- & 0.0033\\
$A_1^{++\star\star}$      & 1.92(13)  & 0.85  & 0.93 & 0.9964 & -- & 0.0036\\
\hline
\hline
$A_1^{++}$                & 1.493(71) & \bf{0.42} & 0.35 & -- & 0.8056 & 0.1944\\
\hline
\hline
\end{tabular}
\caption{Masses in the $A_1^{++}$ channel at the parameters shown in the header. A wide range of different
  variational sets is employed and the results are compared. In
  particular, we want to highlight the mixing with multi--glueballs
  and bi--torelon states.}
}
\TABLE{
\label{tab:a1pp-4-200}
\begin{tabular}[c]{|c|| c c c c c c |}
\hline
\multicolumn{7}{|c|}{SU(4) at $\beta=10.789$ and $N_L=12$}\\
\hline
\hline
$R^{PC}$ &  $am(\sigma)$ & $|c_n|^2$ & $\chi^2$ & mix$_G$ & mix$_S$ & mix$_T$ \\
\hline
\hline
$A_1^{++}$                &   0.808(15)  &  0.99  &   0.16&0.9916  &  0.0066  &  0.0018\\
$A_1^{++\star}$           &   1.388(50)  &  0.94  &   0.06&0.8399  &  0.1232  &  0.0369\\
$A_1^{++\star\star}$      &   1.605(93)  &  0.93  &   0.35&0.5711 &   \bf{0.1604}  &  \bf{0.2685}\\
$A_1^{++\star\star\star}$ &   1.607(91)  &  0.88  &   0.01&0.5590 &   \bf{0.4122} &   0.0288\\
\hline
\hline
$A_1^{++}$                &0.808(15)  &  0.99  &   0.15&0.9934  &  0.0066    &--\\
$A_1^{++\star}$           &1.400(51)  &  0.94  &   0.04&0.8873   & 0.1127    &--\\
$A_1^{++\star\star}$      &1.596(87)  &  0.88  &   0.13&0.6383  &  \bf{0.3617}    &--\\
$A_1^{++\star\star\star}$ &1.92(16)  &  0.86  &   0.32&0.9878 &   0.0122    &--\\
\hline
\hline
$A_1^{++}$                &0.810(15)  &  0.99  &   0.20&0.9973    &--&    0.0027\\
$A_1^{++\star}$           &1.404(53)  &  0.94  &   0.28&0.9553    &--&    0.0447\\
$A_1^{++\star\star}$      &1.614(90)  &  0.93  &   0.02& 0.6587    &--&    \bf{0.3413}\\
$A_1^{++\star\star\star}$ &1.96(18)  &  0.85  &   0.03&0.9979    &--&    0.0021\\
\hline
\hline
$A_1^{++}$                &1.642(92)  &  0.92  &   0.91&--&    0.8334  &  0.1666\\
\hline
\hline
\end{tabular}
\caption{Masses in the $A_1^{++}$ channel at the parameters shown in the header. A wide range of different
  variational sets is employed and the results are compared. In
  particular, we want to highlight the mixing with multi--glueballs
  and bi--torelon states.}
}
\TABLE{
\label{tab:a1pp-5-200}
\begin{tabular}[c]{|c|| c c c c c c |}
\hline
\multicolumn{7}{|c|}{SU(5) at $\beta=17.107$ and $N_L=12$}\\
\hline
\hline
$R^{PC}$ &  $am(\sigma)$ & $|c_n|^2$ & $\chi^2$ & mix$_G$ & mix$_S$ & mix$_T$ \\
\hline
\hline
$A_1^{++}$                & 0.778(14)  & 0.95  & 0.01 & 0.9914 & 0.0059 & 0.0027\\
$A_1^{++\star}$           & 1.409(56)  & 0.93  & 0.36 & 0.9643 & 0.0267 & 0.0089\\
$A_1^{++\star\star}$      & 1.565(80)  & 0.86  & 0.41 & 0.3586 & \bf{0.5395} & \bf{0.1019}\\
$A_1^{++\star\star\star}$ & 1.606(81)  & 0.88  & 0.94 & 0.5318 & \bf{0.2186} & \bf{0.2496}\\
$A_1^{++\star\star\star\star}$ & 1.97(16) &0.99 & 0.30 &0.9668  &  0.0166 &0.0166\\
\hline
\hline
$A_1^{++}$                &0.778(14)  & 0.95  & 0.01 & 0.9940  & 0.0060& -- \\
$A_1^{++\star}$           &1.402(55)  & 0.92  & 0.22 & 0.9752  & 0.0248& -- \\
$A_1^{++\star\star}$      &1.525(81)  & \bf{0.83}  & 0.22 & 0.3203  & \bf{0.6797}& -- \\
$A_1^{++\star\star\star}$ &1.92(15)  & 0.97  & 0.01 & 0.9830  & 0.0170& -- \\
\hline
\hline
$A_1^{++}$                &0.778(14)  & 0.95 & 0.01 & 0.9973 & -- &0.0027\\
$A_1^{++\star}$           &1.400(53)  & 0.92 & 0.40 & 0.9876 & -- &0.0124\\
$A_1^{++\star\star}$      &1.637(85)  & 0.91 & \bf{1.27} & 0.6247 & -- &\bf{0.3753}\\
$A_1^{++\star\star\star}$ &1.95(15)  & 0.97 & 0.10 & 0.9817 & -- &0.0183\\
\hline
\hline
$A_1^{++}$                &1.565(84)&0.87& 0.24 &--&0.9558 &0.0442\\
\hline
\hline
\end{tabular}
\caption{Masses in the $A_1^{++}$ channel at the parameters shown in the header. A wide range of different
  variational sets is employed and the results are compared. In
  particular, we want to highlight the mixing with multi--glueballs
  and bi--torelon states.}
}
\TABLE{
\label{tab:a1pp-6-200}
\begin{tabular}[c]{|c|| c c c c c c |}
\hline
\multicolumn{7}{|c|}{SU(6) at $\beta=24.845$ and $N_L=12$}\\
\hline
\hline
$R^{PC}$ &  $am(\sigma)$ & $|c_n|^2$ & $\chi^2$ & mix$_G$ & mix$_S$ & mix$_T$ \\
\hline
\hline
$A_1^{++}$                &  0.786(14)  & 0.97 & 0.51 & 0.9935 & 0.0029 & 0.0036 \\
$A_1^{++\star}$           &  1.397(48)  & 0.94 & 0.57 & 0.9529 & 0.0367 & 0.0103 \\
$A_1^{++\star\star}$      &  1.632(84)  & 0.91 & 0.39 & 0.6215 & \bf{0.3170} & 0.0615 \\
$A_1^{++\star\star\star}$ &  1.650(95)  & 0.90 & 0.46 & 0.7162 & 0.0988 & \bf{0.1850} \\
\hline
\hline
$A_1^{++}$                &0.786(14)  & 0.97 & 0.51 & 0.9971 & 0.0029 & -- \\
$A_1^{++\star}$           &1.408(49)  & 0.94 & 0.26 & 0.9653 & 0.0347 & -- \\
$A_1^{++\star\star}$      &1.579(87)  & 0.86 & 0.02 & 0.6877 & \bf{0.3123} & -- \\
\hline
\hline
$A_1^{++}$                &0.789(14) & 0.97 & 0.51 & 0.9963 & -- & 0.0037 \\
$A_1^{++\star}$           &1.392(50) & 0.92 & 0.41 & 0.9886 & -- & 0.0114 \\
$A_1^{++\star\star}$      &1.71(11) & 0.96 & \bf{1.43} & 0.6609 & -- & \bf{0.3391} \\
\hline
\hline
$A_1^{++}$                &1.615(89)  &0.90  &0.02 & -- & 0.9608  &0.0392\\
\hline
\hline
\end{tabular}
\caption{Masses in the $A_1^{++}$ channel at the parameters shown in the header. A wide range of different
  variational sets is employed and the results are compared. In
  particular, we want to highlight the mixing with multi--glueballs
  and bi--torelon states.}
}
\TABLE{
\label{tab:a1pp-7-200}
\begin{tabular}[c]{|c|| c c c c c c |}
\hline
\multicolumn{7}{|c|}{SU(7) at $\beta=33.995$ and $N_L=12$}\\
\hline
\hline
$R^{PC}$ &  $am(\sigma)$ & $|c_n|^2$ & $\chi^2$ & mix$_G$ & mix$_S$ & mix$_T$ \\
\hline
\hline
$A_1^{++}$                & 0.771(15) & 0.95 & 0.33 & 0.9922 & 0.0048 & 0.0029\\
$A_1^{++\star}$           & 1.553(63) & 1.06 & 0.13 & 0.9525 & 0.0423 & 0.0052\\
$A_1^{++\star\star}$      & 1.658(96) & 0.94 & 0.62 & 0.5173 & \bf{0.4752} & 0.0074\\
$A_1^{++\star\star\star}$ & 1.82(11) & 1.01 & 0.01 & 0.6839 & 0.0699 & \bf{0.2461}\\
$A_1^{++\star\star\star\star}$ & 2.05(17) & 1.09 & 0.01 & 0.9720 & 0.0229 & 0.0052\\
\hline
\hline
$A_1^{++}$                & 0.771(14) & 0.95 & 0.36 & 0.9951 & 0.0049 & -- \\
$A_1^{++\star}$           & 1.552(63) & 1.06 & 0.08 & 0.9589 & 0.0411 & -- \\
$A_1^{++\star\star}$      & 1.651(95) & 0.93 & 0.29 & 0.4934 & \bf{0.5066} & -- \\
$A_1^{++\star\star\star}$ & 2.02(17) & 1.06 & 0.12 & 0.9776 & 0.0224 & -- \\
\hline
\hline
$A_1^{++}$                & 0.771(15) & 0.95 & 0.31 & 0.9970 & -- & 0.0030\\
$A_1^{++\star}$           & 1.562(63) & 1.06 & 0.18 & 0.9939 & -- & 0.0061\\
$A_1^{++\star\star}$      & 1.84(11) & 1.03 & 0.19 & 0.7515 & -- & \bf{0.2485}\\
$A_1^{++\star\star\star}$ & 2.05(16) & 1.10 & 0.12 & 0.9928 & -- & 0.0072\\
\hline
\hline
$A_1^{++}$                & 1.602(87) & 0.87 & 0.06 & -- & 0.8816 & 0.1184\\
$A_1^{++\star}$           & 2.30(36) & \bf{0.84}& 0.15 & -- & 0.5081 & 0.4919\\
\hline
\hline
\end{tabular}
\caption{Masses in the $A_1^{++}$ channel at the parameters shown in the header. A wide range of different
  variational sets is employed and the results are compared. In
  particular, we want to highlight the mixing with multi--glueballs
  and bi--torelon states.}
}
\clearpage
\TABLE[ht]{
\label{tab:large-a1}
\begin{tabular}[c]{|c|| c c c c c c |}
\hline
\multicolumn{7}{|c|}{Masses in the $A_1$ representation}\\
\hline
\hline
& $A_1^{++}$ & $A_1^{++\star}$ & $A_1^{++}$(S) & $A_1^{++\star\star}$  & $A_1^{-+}$ & $A_1^{-+}$(S) \\
SU(3) & 0.792(14) & 1.370(50) & \bf{1.485(82)}  &  1.86(13) & 1.370(56)    &  -- \\
SU(4) & 0.808(15) & 1.388(50) &   1.642(92)       &  1.92(16) & 1.377(53    & 2.54(52) \\
SU(5) & 0.778(14) & 1.409(56) &   1.565(84)       &  1.97(16) & 1.400(60)    & -- \\
SU(6) & 0.786(14) & 1.397(48) &   1.615(89)       &       --     & \bf{1.315(48)} & 2.43(40) \\
SU(7) & 0.771(15) & 1.553(63) &   1.602(87)       &  2.05(17) & \bf{1.336(51)} & -- \\
SU(8) & 0.785(16) & 1.408(63) &      --               &       --     &        --         & -- \\ 
\hline
SU($\infty$)  & 0.778(8) & 1.456(41) & 1.578(47) & 2.061(28) & 1.407(17) & 2.33 \\ 
SU($\infty$)  & 0.787(5) & 1.412(24) & 1.604(16) & 1.937(41) & 1.381(9)   & 2.47(6) \\ 
\hline
\end{tabular}
\caption{Values of the masses in units of the lattice spacing $a$ for
  each SU($N$) gauge group in the $A_1$ representation. 
  The SU($\infty$) masses comes from a constant fit (lower)
  and a constant plus 1/$N^2$ corrections (upper). Here we quote only
  best $\chi^2$ fits done without using boldface values in the
  table. %The scattering state $A_1^{++}$(S) is extrapolated, but we can
%not distinguish between 1/$N^2$ or 1/N corrections.
}
}
%\begin{minipage}[0]{\textwidth}

\TABLE[ht]{
\label{tab:large-a2}
\begin{tabular}[c]{c|c||ccc|c}

\cline{2-5}
&\multicolumn{4}{|c|}{ Masses in the $A_2$ representation}&\\
\hhline{~:====:~}
\parbox{1.1cm}{~}&&   $A_2^{++}$(2T) &   $A_2^{+-}$ & $A_2^{--}$&\parbox{1.1cm}{~}\\
&  SU(3) &      1.81(12) &   2.33(38)       &  2.41(42) &\\
&  SU(4) & \bf{1.83(15)} &   2.08(19)       &  2.34(37)&\\
&  SU(5) &   1.88(16)    &   2.19(22)       &  -- &\\
&  SU(6) &  --           &   2.59(51)        &  -- &\\
&  SU(7) & \bf{1.84(13)} &  \bf{2.43(43)} &  -- &\\
&  SU(8) &    --         &  \bf{2.21(27)} &  --&\\
\cline{2-5}

& SU($\infty$)  & -- &  2.61(31) & 2.25&\\ 
& SU($\infty$)  & -- &  2.16(9) & 2.37(4)&\\ 
\cline{2-5}
\end{tabular}
%~~~~~~~~~~~~~~~~~
\caption{Values of the masses in units of the lattice spacing $a$ for
  each SU($N$) gauge group in the $A_2$ representation. Results are
  obtained using only single--glueballs and bi--torelons operators. 
  The SU($\infty$) masses comes from a constant fit (lower)
  and a constant plus 1/$N^2$ corrections (upper). Here we quote only
  best $\chi^2$ fits done without using boldface values in the table.}
}
%\TABLE[ht]{
%\label{tab:large-a2}
%%\begin{tabular}[c]{|c||ccc|}
%\begin{tabular}[c]{|c|| p{1.0in} p{1.0in}p{1.0in}p{0in}|}
%\hline
%%\multicolumn{4}{|p{3.20in}|}{\centerline{Masses in the $A_2$ representation}}\\
%\multicolumn{4}{|c}{\centering Masses in the $A_2$ representation}&\\
%\hline
%\hline
%& \centering  $A_2^{++}$(2T) & \centering  $A_2^{+-}$ & \centering$A_2^{--}$&\\
%\centering  SU(3) &\centering      1.81(12) & \centering  2.33(38)       & \centering 2.41(42) &\\
%\centering  SU(4) &\centering \bf{1.83(15)} & \centering  2.08(19)       & \centering 2.34(37)&\\
%\centering  SU(5) &\centering   1.88(16)    & \centering  2.19(22)       & \centering -- &\\
%\centering  SU(6) &\centering  --           & \centering  2.59(51)        &\centering  -- &\\
%\centering  SU(7) &\centering \bf{1.84(13)} & \centering \bf{2.43(43)} & \centering -- &\\
%\centering  SU(8) &\centering    --         &\centering  \bf{2.21(27)} & \centering --&\\
%\hline
%\centering SU($\infty$)  &\centering -- & \centering 2.61(31) &\centering 2.25&\\ 
%\centering SU($\infty$)  &\centering -- & \centering 2.16(9) & \centering2.37(4)&\\ 
%\hline
%\end{tabular}
%%~~~~~~~~~~~~~~~~~
%\caption{Values of the masses in unit of the lattice spacing $a$ for
%  each SU(N)   gauge group in the $A_2$ representation. Results are
%  obtained using only single--glueballs and bi--torelons operators. 
%  The SU($\infty$) masses comes from a constant fit (lower)
%  and a constant plus 1/$N^2$ corrections (upper). Here we quote only
%  best $\chi^2$ fits done without using boldface values in the table.}
%}
%%\end{minipage}

\TABLE{
\label{tab:large-e}
\begin{tabular}[c]{|c|| c c c c c c c |}
\hline
\multicolumn{8}{|c|}{Masses in the $E$ representation}\\
\hline
\hline
& $E^{++}$ & $E^{++\star}$& $E^{++}$(2T) & $E^{+-}$ & $E^{-+}$ & $E^{-+\star}$ & $E^{--}$\\
SU(3) & 1.282(38) & --         & 1.601(82)        &    2.57(44) &     1.657(95) & 2.43(38) & -- \\
SU(4) & 1.290(40) & --         & 1.617(79)        & \bf{2.85(70)} & 1.683(93) & --          & 2.25(25) \\
SU(5) & 1.315(44) & 1.91(14)   & 1.591(80)        &    2.85(91) &     1.73(10) & 2.48(43)   & 2.31(30) \\
SU(6) & 1.293(43) & 1.82(12)   & \bf{1.620(87)} &    2.80(82) &     1.78(11) & 2.56(45)   & 2.20(24) \\
SU(7) & 1.279(44) & 1.99(15)   & 1.82(11)          &    2.58(62) &     1.78(11) &  2.59(42)   & \bf{2.16(28)} \\
SU(8) & 1.339(53) & 1.80(14)   & 1.78(12)          &    2.67(72) &     1.90(14) &     --         & 2.50(41) \\
\hline
SU($\infty$) & 1.310(15) & 1.82(16) & -- & 2.714(86) & 1.830(35) & 2.589(37) & 2.35(15) \\
SU($\infty$) & 1.296(8) & 1.871(42) & -- & 2.64(5) & 1.738(32) & 2.569(37) & 2.777(52) \\
\hline
\end{tabular}
\caption{Values of the masses in units of the lattice spacing $a$ for
  each SU($N$) gauge group in the $E$ representation. 
 The SU($\infty$) masses comes from a constant fit (lower)
and a constant plus 1/$N^2$ corrections (upper). Here we quote only
best $\chi^2$ fits done without using boldface values in the table.}
}

\TABLE[ht]{
\label{tab:large-t1}
\begin{tabular}[c]{|c|| c c c c c|}
\hline
\multicolumn{6}{|c|}{Masses in the $T_1$ representation}\\
\hline
\hline
& $T_1^{++}$ & $T_1^{+-}$ & $T_1^{+-\star}$ &$T_1^{-+}$ & $T_1^{--}$(2T)\\
SU(3) & 2.05(21) & 1.611(78) &      --      & -- & 2.35(33) \\
SU(4) & 2.13(22) & 1.639(77) &  2.15(19) & 2.37(36) & 2.45(39) \\
SU(5) & 2.09(21) & 1.625(76) &  1.99(16) & 2.30(30) & \bf{ 2.16(28) } \\
SU(6) & 2.30(28) & 1.658(77) &  1.94(16) & 2.43(43) & -- \\
SU(7) & 2.42(31) & 1.617(82) &  1.89(15) & \bf{2.48(46)} & -- \\
SU(8) & 2.25(33) & 1.693(97) &  1.98(17) & 2.57(57) & 2.16(27) \\
\hline
SU($\infty$) & 2.310(80) & 1.659(19) & 1.840(55) & 2.50(14) & -- \\
SU($\infty$) & 2.165(54) & 1.638(11) & 1.980(40) & 2.378(48) & -- \\
\hline
\end{tabular}
\caption{Values of the masses in units of the lattice spacing $a$ for
  each SU($N$) gauge group in the $T_1$ representation. Results are
  obtained using only single--glueballs and bi--torelons operators.
 The SU($\infty$) masses comes from a constant fit (lower)
and a constant plus 1/$N^2$ corrections (upper). Here we quote only
best $\chi^2$ fits done without using boldface values in the table.}
}

\TABLE[ht]{
\label{tab:large-t2}
\begin{tabular}[c]{|c|| c c c c c c|}
\hline
\multicolumn{7}{|c|}{Masses in the $T_2$ representation}\\
\hline
\hline
& $T_2^{++}$ & $T_2^{++\star}$ & $T_2^{+-}$ & $T_2^{+-}$(2T) & $T_2^{-+}$ & $T_2^{--}$\\
SU(3) & 1.292(41) & 1.79(11)         & -- & 1.87(14) & 1.665(93) & 2.26(28) \\
SU(4) & 1.333(47) & 1.76(11)         & -- & 1.96(16) & 1.678(95) &  \bf{2.21(25)} \\
SU(5) & 1.391(48) & 1.91(14)         & \bf{1.90(15)} & 1.94(15) & 1.71(10) & \bf{2.01(21)} \\
SU(6) & 1.263(39) & 1.99(14)         & 2.08(18) & -- & 1.85(12) & 2.31(30) \\
SU(7) & 1.367(50) & 1.96(15)         & \bf{2.07(17)} & 2.23(23) & 1.77(11) & 2.45(33) \\
SU(8) & 1.393(57) & \bf{1.90(15)} & 2.07(21) & 2.26(28) & 1.79(14) & 2.20(32) \\
\hline
SU($\infty$) & 1.354(42) & 1.983(71) & 2.05 & -- & 1.812(38) & 2.327(91) \\
SU($\infty$) & 1.329(22) & 1.857(39) & 2.076(5) & -- & 1.73(3) & 2.299(51) \\
\hline
\end{tabular}
\caption{Values of the masses in units of the lattice spacing $a$ for
  each SU($N$) gauge group in the $T_2$ representation. Results are
  obtained using only single--glueballs and bi--torelons operators.
 The SU($\infty$) masses comes from a constant fit (lower)
and a constant plus 1/$N^2$ corrections (upper). Here we quote only
best $\chi^2$ fits done without using boldface values in the table.}
}
\TABLE[ht]{
\label{tab:su-inf}
\begin{tabular}[c]{|c|| c c c c|}
\hline
\multicolumn{5}{|c|}{SU($\infty$)}\\
\hline
\hline            
$R^{PC}$ & $am(\sigma)$ & c & range N  & $\tilde{\chi}^2$ \\
\hline            
$A_1^{++}$ & 0.778(8) & 0.18(0.15) & (3,4,5,6,7,8) & 0.66\\
%&0.766(11)&0.57(0.28)&(4,5,6,7,8)&0.49\\
%&0.781(13)&-0.04(0.46)&(5,6,7,8)&0.33\\
$A_1^{++\star}$ & 1.456(41) & -0.87(0.68) & (3,4,5,6,7,8) & 1.08\\
%&1.477(66)&-1.5(1.7)&(4,5,6,7,8)&1.35\\
$A_1^{++\star\star}$ & 2.061(28) & -1.9(0.4) & (3,4,5,7) & 0.03\\
$A_1^{++}$(S) & 1.578(47) & 0.71(1.1) & (4,5,6,7) & 0.17\\
&1.643(40)&-1.2(0.6)&(3,4,5,6,7)&0.30\\
$A_1^{-+}$ & 1.407(17) & -0.36(0.22) & (3,4,5) & 0.04\\
& 1.331(28) & 0.47(0.47) & (3,4,5,6,7) & 0.41\\
$A_1^{-+}$(S) & 2.33 & 3 & (4,6) & --\\ 
\hline            
$A_2^{+-}$ &  2.61(31) & -9(6) & (4,5,6) & 0.29\\
& 2.26(14) & -1.1(2.5) & (3,4,5,6,7,8) & 0.34\\
$A_2^{--}$ & 2.25 & 1.4 & (3,4) & --\\
\hline            
$E^{++}$ & 1.310(15) & -0.26(0.24) & (3,4,5,6,7,8) & 0.23\\
$E^{++\star}$ & 1.82(16) & 2(6) & (5,6,7,8) & 0.57\\
$E^{+-}$ & 2.714(86) & -1.2(1.2) & (3,5,6,7,8) & 0.03\\
& 2.742(99) & -1.2(1.4) & (3,4,5,6,7,8) & 0.04\\
$E^{-+}$ & 1.830(35) & -1.8(0.6) & (3,4,5,6,7,8) & 0.20\\
%& 1.882(39)&-3(1) & (4,5,6,7,8)& 0.12\\
$E^{-+\star}$ & 2.589(37) & -1.5(0.6) & (3,5,6,7) & 0.01\\
$E^{+-}$ & 2.714(86) & -1.2(1.2) & (3,5,6,7,8) & 0.03\\
& 2.742(99) & -1.2(1.4) & (3,4,5,6,7,8) & 0.04\\
$E^{--}$ & 2.35(15)&-2(4)&(4,5,6,8)&0.18\\
&2.26(13)&-0.2(3.2)&(4,5,6,7,8)&0.18\\
\hline            
$T_1^{++}$ & 2.310(80) & -2.5(1.2) & (3,4,5,6,7,8) & 0.17\\
$T_1^{+-}$ & 1.659(19) & -0.4(0.3) & (3,4,5,6,7,8) & 0.11\\
& 1.621(51) & 0.1(0.8) & (3,4,5,6,7,8) & 0.79\\
$T_1^{+-\star}$ & 1.840(55) & 4(2) & (4,5,6,7,8) & 0.11\\
$T_1^{-+}$ & 2.50(14) & -3(3) & (4,5,6,8) & 0.06\\
& 2.524(97) & -3(2) & (4,5,6,7,8) & 0.04\\
\hline            
$T_2^{++}$ & 1.354(42) & -0.5(0.7) & (3,4,5,6,7,8) & 1.58\\
& 1.309(22) & -0.07(0.38) & (3,4,5,6,7,8) & 0.50\\
$T_2^{++\star}$ & 1.983(71) & -2(1) & (3,4,5,6,7) & 0.38\\
& 1.965(54) & -1.9(0.8) & (3,4,5,6,7,8) & 0.30\\
$T_2^{+-}$ & 2.05 & 0.8 & (6,8) & --\\
& 2.23(8) & -8(3) & (5,6,7,8) & 0.09\\
$T_2^{-+}$ & 1.812(38) & -1.5(0.6) & (3,4,5,6,7,8) & 0.23\\
$T_2^{--}$ & 2.327(91) & -0.6(1.5) & (3,6,7,8) & 0.15\\
& 2.21(13) & -0.1(2.2) & (3,4,5,6,7,8) & 0.39\\
\hline
\hline
\end{tabular}
\caption{Spectrum of the SU($\infty$) lattice gauge theory. The
  masses, in units of lattice spacing, are obtained from fits over the
  range of $N$ shown; also the fitted
  coefficient $c$ of the $1/N^2$ correction is shown. When 2 fits are
  shown, the second one includes masses that we defined not reliable
  if the range of N is different.}
}

%----------------------------SPECTRUM OF A1++ GIVEN SU(N)------------------------%
\FIGURE[htb]{
\epsfig{figure=FIGS/A1pp_SU3_cfr_opbasis_200_boxes.eps,width=0.75\textwidth}
\label{fig:a1pp-3-200}
\caption{Variational calculation for
SU(3) using different sets of operators on a $N_L=12$ lattice. The unfilled symbols
represent masses that cannot be reliably interpreted as pure glueballs.}
}
% \FIGURE[ht]{
% \epsfig{figure=FIGS/A1pp_SU3_cfr_opbasis_250_boxes.eps,width=0.75\textwidth}
% \label{fig:a1pp-3-250}
% \caption{$N_{compound}=250$ and $N_L=12$. Variational calculation for SU(3) using different sets of operators. The unfilled symbols represents masses that we cannot reliably interpret as pure glueballs. }
% }
% \FIGURE[ht]{
% \epsfig{figure=FIGS/A1pp_SU3_L18_cfr_opbasis_250_boxes.eps,width=0.75\textwidth}
% \label{fig:a1pp-3-18-250}
% \caption{$N_{compound}=250$ and $N_L=18$. Variational calculation for SU(3) using different sets of operators. The unfilled symbols represents masses that we cannot reliably interpret as pure glueballs. }
% }
\FIGURE[ht]{
\epsfig{figure=FIGS/A1pp_SU4_cfr_opbasis_200_boxes.eps,width=0.75\textwidth}
\label{fig:a1pp-4-200}
\caption{Variational calculation for
SU(4) using different sets of operators on a $N_L=12$ lattice. The unfilled symbols
represent masses that cannot be reliably interpreted as pure glueballs.}
}
% \FIGURE[ht]{
% \epsfig{figure=FIGS/A1pp_SU4_cfr_opbasis_250_boxes.eps,width=0.75\textwidth}
% \label{fig:a1pp-4-250}
% \caption{$N_{compound}=250$ and $N_L=12$. Variational calculation for SU(4) using different sets of operators.  The unfilled symbols represents masses that we cannot reliably interpret as pure glueballs}
% }
\FIGURE[ht]{
\epsfig{figure=FIGS/A1pp_SU5_cfr_opbasis_200_boxes.eps,width=0.75\textwidth}
\label{fig:a1pp-5-200}
\caption{Variational calculation for
SU(5) using different sets of operators on a $N_L=12$ lattice. The unfilled symbols
represent masses that cannot be reliably interpreted as pure glueballs.}
}
% \FIGURE[ht]{
% \epsfig{figure=FIGS/A1pp_SU5_cfr_opbasis_250_boxes.eps,width=0.75\textwidth}
% \label{fig:a1pp-5-250}
% \caption{$N_{compound}=250$ and $N_L=12$. Variational calculation for SU(5) using different sets of operators.  The unfilled symbols represents masses that we cannot reliably interpret as pure glueballs}
% }
\FIGURE[ht]{
\epsfig{figure=FIGS/A1pp_SU6_cfr_opbasis_200_boxes.eps,width=0.75\textwidth}
\label{fig:a1pp-6-200}
\caption{Variational calculation for
SU(6) using different sets of operators on a $N_L=12$ lattice. The unfilled symbols
represent masses that cannot be reliably interpreted as pure glueballs.}
}
% \FIGURE[ht]{
% \epsfig{figure=FIGS/A1pp_SU6_cfr_opbasis_250_boxes.eps,width=0.75\textwidth}
% \label{fig:a1pp-5-250}
% \caption{$N_{compound}=250$ and $N_L=12$. Variational calculation for SU(6) using different sets of operators.  The unfilled symbols represents masses that we cannot reliably interpret as pure glueballs}
% }
\FIGURE[ht]{
\epsfig{figure=FIGS/A1pp_SU7_cfr_opbasis_200_boxes.eps,width=0.75\textwidth}
\label{fig:a1pp-7-200}
\caption{Variational calculation for
SU(7) using different sets of operators on a $N_L=12$ lattice. The unfilled symbols
represent masses that cannot be reliably interpreted as pure glueballs.}
}
\FIGURE[ht]{
\epsfig{figure=FIGS/A1pp_SU3_L18_cfr_opbasis_200_boxes.eps,width=0.75\textwidth}
\label{fig:a1pp-3-18-200}
\caption{Variational calculation for
SU(3) using different sets of operators on a $N_L=18$ lattice. The unfilled symbols
represent masses that cannot be reliably interpreted as pure glueballs.}
}

%--------------------------FIGURES FOR THE LARGE-N EXTRAPOLATION------------------
\clearpage
\FIGURE[ht]{
\epsfig{figure=FIGS/A1pp_scatt_fit.eps,width=0.77\textwidth}
\label{fig:a1pp_ln}
\caption{Extrapolation to $N \to \infty$ of the states in the
$A_1^{++}$ channel.}
}

\FIGURE[ht]{
\epsfig{figure=FIGS/A1mp_scatt_fit.eps,width=0.77\textwidth}
\label{fig:a1mp_ln}
\caption{Extrapolation to $N \to \infty$ of the states in the $A_1^{-+}$ channel. We denote the less reliable states with open symbols.}
}

\FIGURE[ht]{
\epsfig{figure=FIGS/A2pm0_noscatt_fit.eps,width=0.77\textwidth}
\label{fig:a2pm_ln}
\caption{Extrapolation to $N \to \infty$ of the states in the $A_2^{+-}$ channel. We denote the less reliable states with open symbols.}
}

\FIGURE[ht]{
\epsfig{figure=FIGS/Epp_scatt_fit.eps,width=0.77\textwidth}
\label{fig:epp_ln}
\caption{Extrapolation to $N \to \infty$ of the states in the $E^{++}$ channel.}
}

\FIGURE[ht]{
\epsfig{figure=FIGS/Emp_scatt_fit.eps,width=0.77\textwidth}
\label{fig:emp_ln}
\caption{Extrapolation to $N \to \infty$ of the states in the $E^{-+}$ channel.}
}

\FIGURE[ht]{
\epsfig{figure=FIGS/Epm-mm_scatt_fit.eps,width=0.77\textwidth}
\label{fig:epm-mm_ln}
\caption{Extrapolation to $N \to \infty$ of the states in the $E^{+-}$ and $E^{--}$ channels. We denote the less reliable states with open symbols.}
}

\FIGURE[ht]{
\epsfig{figure=FIGS/T1pp-mp_scatt_fit.eps,width=0.77\textwidth}
\label{fig:t1pp-mp_ln}
\caption{Extrapolation to $N \to \infty$ of the states in the $T_1^{++}$ channel and in the $T_1^{-+}$ channel. We denote the less reliable states with open symbols.}
}

\FIGURE[ht]{
\epsfig{figure=FIGS/T1pm_scatt_fit.eps,width=0.77\textwidth}
\label{fig:t1pm_ln}
\caption{Extrapolation to $N \to \infty$ of the states in the $T_1^{+-}$ channel.}
}

\FIGURE[ht]{
\epsfig{figure=FIGS/T2pp_scatt_fit.eps,width=0.77\textwidth}
\label{fig:t2pp_ln}
\caption{Extrapolation to $N \to \infty$ of the states in the $T_2^{++}$ channel. We denote the less reliable states with open symbols.}
}
\FIGURE[ht]{
\epsfig{figure=FIGS/T2pm-mp-mm_scatt_fit.eps,width=0.77\textwidth}
\label{fig:t2pm-mp-mm_ln}
\caption{Extrapolation to $N \to \infty$ of the states in the $T_2^{+-}$, $T_2^{-+}$ and $T_2^{--}$ channels. We denote the less reliable states with open symbols.}
}

\clearpage
%-----------------------SPECTRUM AT INFINITY--------------
\FIGURE[ht]{
\epsfig{figure=FIGS/spectrum_suINF_new.eps,width=0.77\textwidth}
\label{fig:spectrum_ln}
\caption{The spectrum at $N=\infty$. The yellow boxes represent the
large $N$ extrapolation of masses obtained in ref.~\cite{Lucini:2004my}.}
}
%-----------------------REGGE TRAJECTORIES-----------------
\FIGURE[ht]{
\epsfig{figure=FIGS/regge_new.eps,width=0.77\textwidth}
\label{fig:regge_new}
\caption{Chew-Frautschi plot of the glueball spectrum}
}

\end{document}